\documentclass[aps,prl,reprint,superscriptaddress]{revtex4-1}

\usepackage{graphicx}
\usepackage{bm}
\usepackage{amsmath}
\usepackage{comment}

\begin{document}


\title{Optical switching of resonance fluorescence from a single germanium vacancy color center in diamond}


\author{Disheng Chen}
\thanks{These two authors contributed equally}
\affiliation{Division of Physics and Applied Physics, School of Physical and Mathematical Sciences, Nanyang Technological University, Singapore 637371, Singapore}
\affiliation{The Photonics Institute and Centre for Disruptive Photonic Technologies, Nanyang Technological University, Singapore 637371, Singapore}

\author{Zhao Mu}
\thanks{These two authors contributed equally}
\affiliation{Division of Physics and Applied Physics, School of Physical and Mathematical Sciences, Nanyang Technological University, Singapore 637371, Singapore}


\author{Yu Zhou}
\affiliation{Division of Physics and Applied Physics, School of Physical and Mathematical Sciences, Nanyang Technological University, Singapore 637371, Singapore}

\author{Johannes Froech}
\affiliation{School of Mathematical and Physical Sciences, University of Technology Sydney, Ultimo, NSW, 2007, Australia}

\author{Carole Diederichs}
\affiliation{MajuLab, International Joint Research Unit UMI 3654, CNRS, Université Côte d’Azur, \\ Sorbonne Université, National University of Singapore, Nanyang Technological University, Singapore}

\author{Nikolay Zheludev}
\affiliation{Division of Physics and Applied Physics, School of Physical and Mathematical Sciences, Nanyang Technological University, Singapore 637371, Singapore}
\affiliation{The Photonics Institute and Centre for Disruptive Photonic Technologies, Nanyang Technological University, Singapore 637371, Singapore}
\affiliation{Optoelectronics Research Centre, University of Southampton, UK}

\author{Igor Aharonovich}
\affiliation{School of Mathematical and Physical Sciences, University of Technology Sydney, Ultimo, NSW, 2007, Australia}

\author{Wei-bo Gao}
\affiliation{Division of Physics and Applied Physics, School of Physical and Mathematical Sciences, Nanyang Technological University, Singapore 637371, Singapore}
\affiliation{The Photonics Institute and Centre for Disruptive Photonic Technologies, Nanyang Technological University, Singapore 637371, Singapore}


\date{\today}

\begin{abstract}
Scalable quantum photonic networks require coherent excitation of quantum emitters. However, many solid-state systems can undergo a transition to a dark shelving state that inhibits the fluorescence. Here we demonstrate that a controlled gating using a weak non-resonant laser, the resonant excitation can be recovered and amplified for single germanium vacancies (GeVs).  Employing the gated resonance excitation, we achieve optically stable resonance fluorescence of GeV centers. Our results are pivotal for the deployment of diamond color centers as reliable building blocks for scalable solid state quantum networks.
\end{abstract}




\maketitle


Artificial atomic systems that can be coherently controlled and manipulated are of a paramount importance for realization of scalable quantum photonic architectures \cite{aharonovich_diamond_2014, atature_material_2018}. Recently, color centers in diamond, particularly group IV defects, such as the silicon vacancies (SiV) \cite{neu_single_2011} or the germanium vacancies (GeV) \cite{iwasaki_germanium-vacancy_2015, palyanov_germanium:_2015, hausler_photoluminescence_2017, bhaskar_quantum_2017, siampour_-chip_2018, bray_single_2018} have emerged as attractive candidates. These defects possess an inversion symmetry \cite{hepp_electronic_2014} and therefore are not sensitive to local fluctuation in electric fields, resulting in a robust optical fluorescence with high indistinguishability \cite{sipahigil_indistinguishable_2014}. Additional advantage of those systems is their high Debye Waller factor that is manifested in a significant portion of the emission being concentrated in the zero phonon line (ZPL) \cite{neu_fluorescence_2011, palyanov_germanium:_2015}. This high concentration makes their resonance fluorescence (RF) appealing for efficient long-distance quantum communication \cite{duan_long-distance_2001}, quantum teleportation \cite{bouwmeester_experimental_1997} and entanglement swapping \cite{pan_experimental_1998}.

Unfortunately, under resonant excitation, these systems can undergo a non-radiative transition to a dark state, resulting a quenching of RF. For the nitrogen vacancy (NV) centers \cite{doherty_nitrogen-vacancy_2013}, this is often associated with a charge-state transition from negative to neutral \cite{waldherr_dark_2011, siyushev_optically_2013}. Such a process results in lack of photons under resonant excitation, and consequently hinder the potential for single shot spin readout \cite{sukachev_silicon-vacancy_2017, vamivakas_observation_2010}, and continuous operation of the quantum network \cite{kimble_quantum_2008}.   
Here we show that the quenching of RF also occurs for GeV color centers. In the positive side, we find that the RF can be reinstated by employing a small amount of non-resonant beam at 532 nm. This laser acts as a gate control over the fluorescence from the emitter, which can be quantitatively modeled by using a 2-level system accompanied by a dark-state.

\begin{figure}[t]  
	\includegraphics{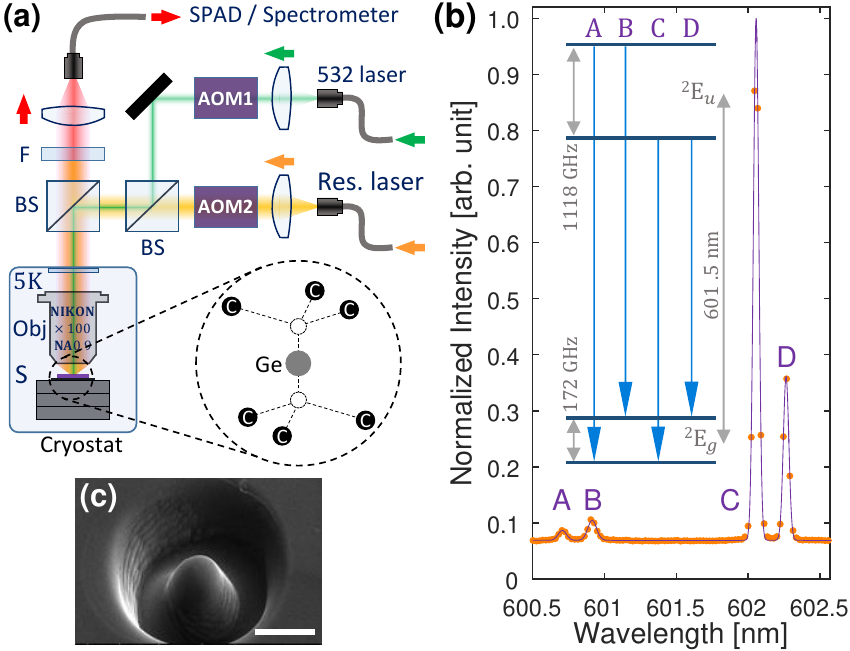}
	\caption{\label{fig:setup}  (a) Experimental setup. AOM: acousto-optic modulator; BS: 50:50 non-polarizing beam splitter; Obj: objective; S: sample; F: band-pass filter; SPAD: single-photon avalanche detector. Bottom: schematic of a GeV center. (b) Normalized PL spectrum of the GeV color center at 5 K, excited at 532 nm with a power of 0.4 mW (0.06 P$_1$) for an exposure time of 5 s. Purple line is the fitting with four Gaussian peaks, labeled as A, B, C, and D from high to low energy. Inset: energy structure of the GeV center with four optical transitions labeled according to the spectrum. Splitting reflects the best-fit parameters. (c) Scanning electron microscope (SEM) image of a FIB milled SIL. Scale-bar:~3~$\mu$m. 
	}
\end{figure}

The investigated sample consists of implantation-generated GeV centers within an electronic-grade Type IIa diamond \cite{GeV_supp_2019}. The implanted Ge atom takes the interstitial space between the two empty carbon sites, forming a unique split-vacancy configuration with D$_{3d}$ symmetry, as shown in Fig.~\ref{fig:setup}(a). Due to the strong spin-orbit coupling \cite{thiering_ab_2018}, the ground state ($^{2}\text{E}_g$) and excited state ($^2\text{E}_u$) split into a pair of energy levels with two-fold spin-degeneracy at zero magnetic field, leading to the characteristic four-line fine structure in the ZPL emission spectrum at 602 nm [Fig.~\ref{fig:setup}(b)]. To enhance the photon collection efficiency, a half-sphere solid immersion lens (SIL) with a diameter of 5 $\mu$m is fabricated on top of the sample by using focused ion (Ga+) beam (FIB) milling before Ge implantation \cite{marseglia_nanofabricated_2011, GeV_supp_2019}, as shown in Fig.~\ref{fig:setup}(c). The sample is mounted on a XYZ piezo-stepper motorized stage housed in a closed-cycle helium-flow cryostat at 5 K.  

All optical measurements are performed by using a home-built confocal microscope, as shown in Fig.~\ref{fig:setup}(a). An achromatic microscopic objective with NA=0.9 is placed one focal length away from the sample to focus the excitation beam into the SIL and collect the PL from the emitter. A tunable continuous-wave (cw) laser with a linewidth of $<1$ MHz is used to resonantly address the GeV center, and perform photoluminescence excitation (PLE) measurements. A diode-pumped solid-state laser at 532 nm is used for non-resonant excitation of the emitter and gating of RF, enabled by passing through an acousto-optic modulator (AOM). After directed through a band-pass filter, the PL is coupled into a single-mode fiber connected to a spectrometer or a single-photon avalanche detector (SPAD). In PL spectrum characterization, a $600\pm7$ nm band-pass filter is used for ZPL detection; in PLE and gating experiments, a $650\pm20$ nm band-pass filter is used for phonon-side band (PSB) PL detection.

\begin{figure}[t]
	\includegraphics{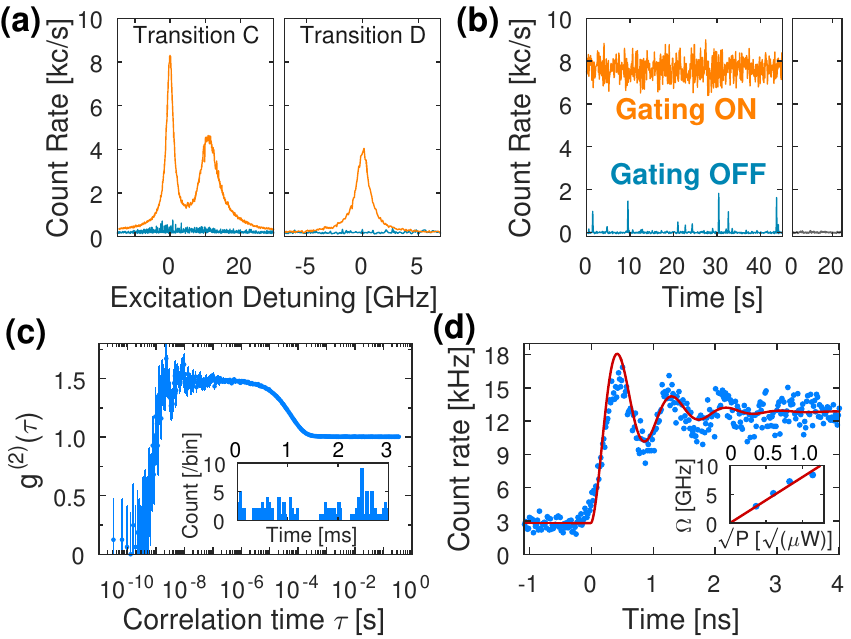}
	\caption{\label{fig:gating} (a) PLE spectra of transitions C (left) and D (right) when the gating laser is on (orange) and off (blue). Zero detuning corresponds to 602.2903 nm and 602.4828 nm for C and D, respectively. (b) RF intensity of transition C for gating on (orange) and gating off (blue). Right: PL intensity under solely non-resonant excitation. Background has been subtracted from the data. Binsize: 100 ms. (c) Second-order correlation function of the GeV center under resonant pumping of transition C with $g^{(2)}(0)=0.07$. Inset: Stochastic jump of the RF. Binsize: 33 $\mu$s. (d) Rabi oscillation of transition C. Red curve is a fitting with 2-level model \cite{GeV_supp_2019}. Inset: Rabi frequency versus square root of resonant power with a linear fit (red). For all data, resonant power is 200 nW (0.35~P$_0$), and non-resonant power is 1.2 $\mu$W (1.8$\times$10$^{-4}$~P$_1$). }
\end{figure}

The gating effect by the non-resonant laser can be demonstrated by comparing PLE spectra with the gating laser on or off, as shown in Fig.~\ref{fig:gating}(a). For both transitions C and D, the PLE spectra are only detectable when the gating laser is on. The multiple peaks around transition C possibly originate from the nearby GeV centers, whose associated D lines are shifted out of the measurement window thanks to the different strains experienced by centers \cite{GeV_supp_2019}. The PL intensity is enhanced by 500 folds when switching on the gating laser, as shown in Fig.~\ref{fig:gating}(b), where the gating power is $\sim$10$^{-4}$ of non-resonant saturation power P$_1 = 6.8 \pm 0.1$ mW \cite{GeV_supp_2019}. In fact, this non-resonant beam is too weak to induce any detectable fluorescence from the emitter [right panel of Fig.~\ref{fig:gating}(b)], and the main role played by this light is a switch controlling the on and off of the RF from the emitter. We stress that the optical pumping between the two ground states cannot account for the observation because the orbital relaxation, T$_1^\text{orbital}\simeq20$ ns \cite{siyushev_optical_2017}, is orders of magnitude faster than the gating dynamics involved here. Instead, a long-lived dark state is resorted for the explanation, evident by the bunching plateau of second-order correlation function and the stochastic jumping of RF, as shown in Fig.~\ref{fig:gating}(c) \cite{nguyen_photoneutralization_2013, delteil_observation_2014}. Even with the presence of dark state, coherence between ground and excited states can still be generated and maintained for a coherence time of T$_2=366\pm20$ ps, as shown by the Rabi oscillation of transition C in Fig.~\ref{fig:gating}(d). Since both transitions C and D are equivalent for our study, we focus on the latter for the rest of the Letter for the sake of clarity.

\begin{figure}[t]
	\includegraphics{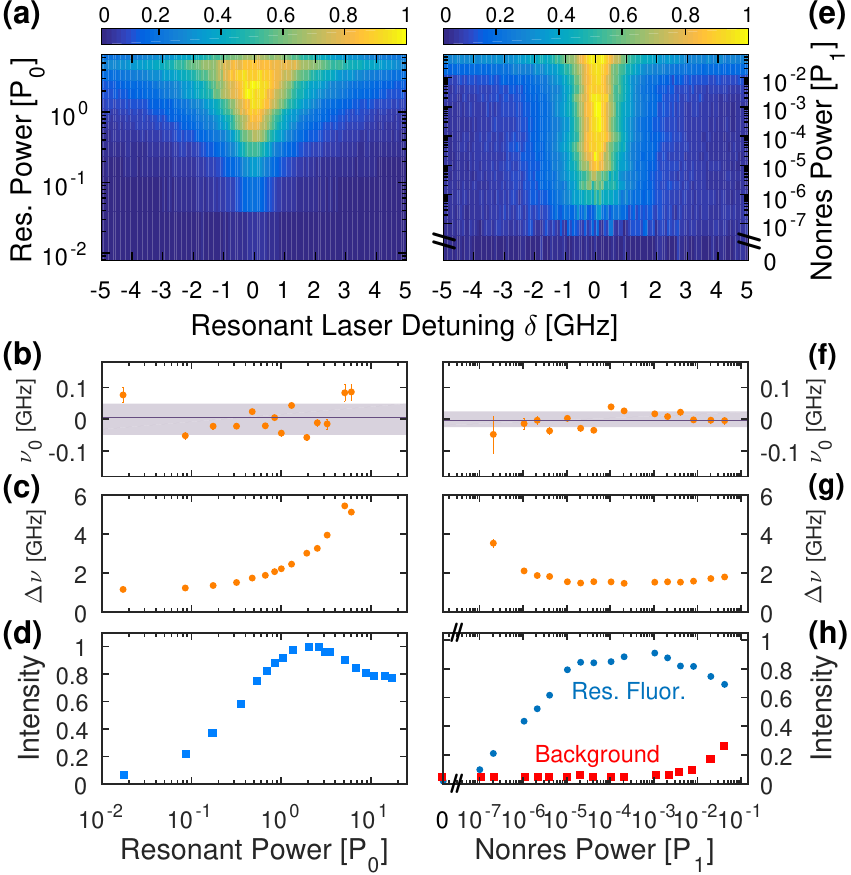}
	\caption{\label{fig:ple}  2D map of normalized PLE spectra (transition D) by varying (a) the resonant excitation power, or (e) the non-resonant excitation power. Normalization constant: (a) 10 kcnt/s, and (e) 4 kcnt/s. Gating power in (a): $7\times10^{-5}$ P$_1$; resonant power in (e): 0.35 P$_0$. (b) and (f) are the center frequency $\nu_0$ of each line in (a) and (e), respectively, extracted from Lorentzian fitting. The shaded region represents the standard deviation of $\nu_0$, (b) $\sigma$$\sim$50 MHz, and (f) $\sigma$$\sim$25 MHz. (c) and (g) are the Lorentzian linewidth $\Delta\nu$ of each line in (a) and (e), respectively. (d) Resonant-power dependence of RF, measured by setting the resonant laser at zero detuning. (h) Gating-power dependence of RF (blue), evaluated by subtracting the background from the maximum count rate of each line in (e). Background count rate (red) is measured at a far-detuning of $\sim$10 GHz. }
\end{figure}

To understand the photodynamics in the system, we study the power dependence of RF by varying either the resonant [Fig.~\ref{fig:ple}(a)] or gating power [Fig.~\ref{fig:ple}(e)]. By fitting each line with a Lorentzian function, we obtain a constant transition energy for different resonant powers [Fig.~\ref{fig:ple}(b)], and observe a pronounced power-broadening [Fig.~\ref{fig:ple}(c)]. Meanwhile, the RF intensity displays an unconventional power dependence characterized by an unexpected drop at $\sim$3 P$_0$, as shown in Fig.~\ref{fig:ple}(d), where P$_0=1.15\pm0.39$ $\mu$W is the resonant saturation power, determined by employing a pulse measurement scheme \cite{GeV_supp_2019}. The drop of RF verifies the existence of dark state, and indicates the opposite role played by the resonant laser to the gating beam, i.e., shelving the population into the dark state. 

As the gating power increases, the initially irresolvable PLE spectrum starts to recover and then stabilizes at $\sim$10$^{-5}$ P$_1$ [Fig.~\ref{fig:ple}(e)]. Through the evolution, the transition shows an exceptional stability by displaying zero drift of transition energy [Fig.~\ref{fig:ple}(f)], and an unvarying excitation linewidth [Fig.~\ref{fig:ple}(g)]. This superior optical property stems from the inversion symmetry of GeV center \cite{siyushev_optical_2017}, and shows a striking contrast to the significant spectral diffusion displayed by NV centers under non-resonant excitation \cite{wolters_measurement_2013}. The slightly broadening of linewidth for the low gating powers ($<10^{-6}$ P$_1$) is caused by the detuning dependence of shelving efficiency. Since the shelving becomes significantly stronger for smaller detuning (given a constant de-shelving rate), it causes a flattening of PLE spectrum, and gives rise to a wider linewidth \cite{GeV_supp_2019}. This is similar to the linewidth broadening observed in SiV center at milli-kelvin temperature, where spin pumping plays the role of shelving \cite{becker_all-optical_2018}. As the gating power increases, the gating-based dynamics is enhanced and finally dominates over the resonant-induced shelving process, thus restoring the linewidth to its intrinsic value. When the gating power exceeds 10$^{-3}$ P$_1$, the RF intensity starts to drop, which is accompanied by a rising of PLE background produced by non-resonant excitation [Fig.\ref{fig:ple}(h)]. This reveals a competition between the resonant and non-resonant excitation.

\begin{figure*}[t]
	\includegraphics{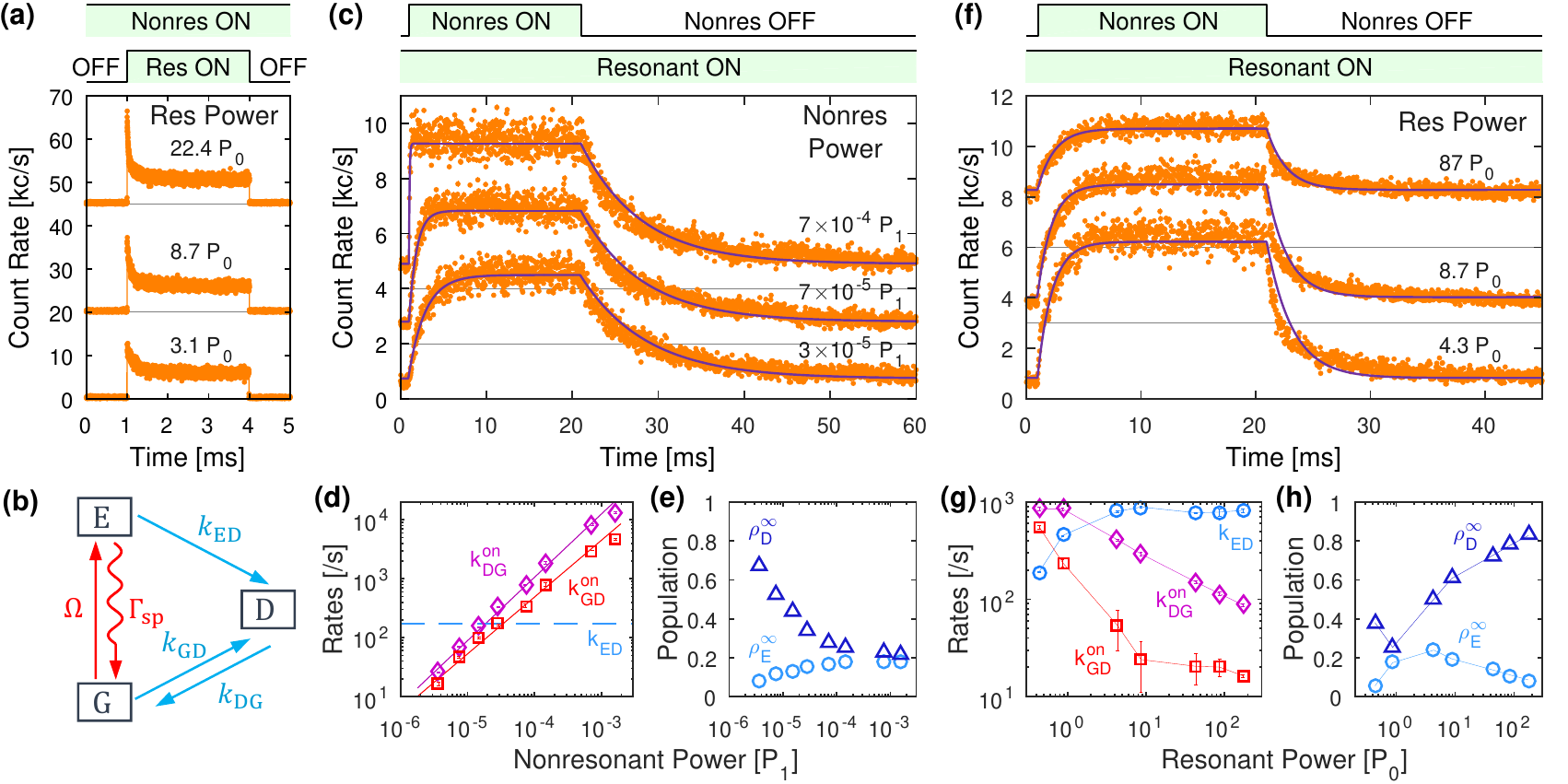}
	\caption{\label{fig:onoff}  Gating and shelving dynamics. (a) Time-resolved PL by modulating the resonant beam with constant non-resonant power of $7\times10^{-7}$ P$_1$. (b) Physical model. G, E and D: ground, excited and dark state;  $k_\text{GD}$, $k_\text{DG}$ and $k_\text{ED}$: population transfer rates from G to D, D to G, and E to D; R: resonant excitation rate; $\Gamma_\text{sp} = 1/\text{T}_1 = 280$ MHz: spontaneous decay rate, determined by lifetime measurement \cite{GeV_supp_2019}. (c) and (f) are the time-resolved PL by modulating the non-resonant beam with (c) constant resonant power (0.9 P$_0$) or (f) constant non-resonant power ($7\times10^{-5}$ P$_1$). Black curves are the fittings by using Eqn.~\ref{eqn:I}. (d) and (g) are the dynamical rates extracted from the fittings in (c) and (f), respectively. Dashed blue horizontal lines in (d) depicts $k_\text{ED}$, representing its trivial non-resonant power dependence in this experiment. Solid straight lines in (d) are the fittings with $k_\text{GD}^\text{on} = 3.5 \times 10^6 \times P^{0.96}$ (red) and $k_\text{DG}^\text{on} = 2.1 \times 10^7 \times P^{1.07}$ (purple), where P denotes the non-resonant power in the unit of P$_1$. (e) and (h) are the on-period steady-state population of dark state $\rho_\text{D}^{\infty}$ and 2-level system $\rho_\text{E}^{\infty}$, evaluated by using the rates in (d) and (g), respectively. In (a), (c) and (f), raw data (orange dots) are vertically shifted for clarity, with the zero-intensity level indicated by the grey horizontal lines. Top panel: modulation protocol. }
\end{figure*}

The shelving effect induced by the resonant laser can be directly observed by modulating the resonant beam while keeping the non-resonant beam in cw-mode, as shown in Fig.~\ref{fig:onoff}(a). The immediate exponential decay of RF following the  excitation edge directly monitors the shelving process. The hight of the transient peak reflects the population in the excited state before it is influenced by the shelving process induced by the resonant pumping. The subsequent plateau corresponds to the equilibrium state of the system dictated by both shelving and de-shelving rates. Following this phenomenological picture, we construct a 3-state model composed of a 2-level system and a dark state, as shown in Fig.~\ref{fig:onoff}(b). The population in the ground state (G) can be resonantly promoted ($\Omega$) to the excited state (E), where the population can either relax back to the ground state via spontaneous decay ($\Gamma_\text{sp}$), or be shelved into a dark state (D) non-radiatively ($k_\text{ED}$) via resonant pumping. The ground and dark state can exchange the population at rates $k_\text{DG}$ and $k_\text{GD}$, mainly enabled by non-resonant pumping. Within the framework of semi-classical picture, the time-evolution of the system follows the master equation

\begin{align}
\label{eqn:rate}
&\frac{d}{dt} \left(\begin{array}{ccccc}
\rho_\text{G}, & \rho_\text{E}, & \rho_\text{GE}, & \rho_\text{EG}, & \rho_\text{D}
\end{array}\right)^\text{T} = \nonumber \\ 
&\left(
\begin{array}{ccccc}
-k_\text{GD} & \Gamma_\text{sp} & i\Omega/2 & -i\Omega/2 & k_\text{DG} \\
0 & -\Gamma_\text{sp}-k_\text{ED} & -i\Omega/2 & i\Omega/2 & 0 \\
i\Omega/2 & -i\Omega/2 & -1/T_2 & 0 & 0 \\
-i\Omega/2 & i\Omega/2 & 0 & -1/T_2 & 0 \\
k_\text{GD} & k_\text{ED} & 0 & 0 & -k_\text{DG}
\end{array}\right) \left( 
\begin{array}{c}
\rho_\text{G}\\
\rho_\text{E}\\
\rho_\text{GE}\\
\rho_\text{EG}\\
\rho_\text{D}
\end{array}\right)
\end{align}
where $\rho_\text{G}$, $\rho_\text{E}$, and $\rho_\text{D}$ are the time-dependent population in ground, excited and dark state, $\rho_\text{GE}$ and $\rho_\text{GE}$ are the coherence between G and E, $\Omega$ is the resonant Rabi frequency, $\Gamma_\text{sp}$ is the spontaneous decay rate, and $\text{T}_2$ is the coherence time of excited state. The excitation linewidth can be derived from the steady-state solution of Eqn.~\ref{eqn:rate} 

\begin{equation}
\label{eqn:linewidth}
\Delta\nu = \frac{1}{\pi T_2} \sqrt{1+\frac{1}{2}\frac{\Omega^2T_2(k_\text{ED}+2k_\text{DG}+k_\text{GD})}{(\Gamma_\text{sp}+k_\text{ED})(k_\text{DG}+k_\text{GD})}}
\end{equation}
in the unit of linear frequency. By equalizing the asymptotic linewidth at 0 P$_0$ in Fig.~\ref{fig:ple}(c) ($\sim$1 GHz, 20 times of lifetime-limited value) to Eqn.~\ref{eqn:linewidth} with $\Omega=0$, we find $\text{T}_2 = 316\pm20$ ps, consistent with the coherence time obtained from Rabi oscillation measurement [Fig.~\ref{fig:gating}(d)]. The detected RF intensity follows
\begin{equation}
\label{eqn:I}
I_\text{PL}(t) = \eta\Gamma_\text{sp}\rho_\text{E}(t)
\end{equation}
where $\eta=9\times10^{-5}$ is the overall efficiency including both detection efficiency of the experimental setup and quantum yield of GeV center \cite{GeV_supp_2019, boldyrev_bright_2018}. 


To extract the dynamical rates of gating and shelving, we perform a similar time-resolved experiment, but modulating the non-resonant beam while keeping the resonant beam in cw mode. Here, the PL inherits the modulation pattern of the gating laser, and displays a gating-power-dependent modulation depth, as shown in Fig.~\ref{fig:onoff}(c). Since the non-resonant laser has little effect on $k_\text{ED}$, we keep this rate a constant and determine it via global fitting  \cite{GeV_supp_2019}. The main effect of the gating beam is to promote $k_\text{GD}$ and $k_\text{DG}$ linearly over the non-resonant power, as show in Fig.~\ref{fig:onoff}(d). This power dependence implies a single-photon process for the shelving and deshelving of population induced by the non-resonant laser. Consequentially, the steady-state population is transfered from the dark state to the ground and excited states as increasing the gating power, as shown in Fig.~\ref{fig:onoff}(e). 


Resonant power dependence is shown in Fig.~\ref{fig:onoff}(f). The main effect of the resonant laser is to speed up the shelving rate $k_\text{ED}$, while indirectly reducing rates $k_\text{GD}$ and $k_\text{DG}$, as shown in Fig.~\ref{fig:onoff}(g). The saturation behavior of $k_\text{ED}$ implies a two-step shelving process mediated by a meta-stable state. The first step of population pumping from the excited state to the meta-stable state is responsible for the enhancement of $k_\text{ED}$, while the second step of non-radiative decay from the metastable state to the dark state caps $k_\text{ED}$ at kHz regime. The peak of steady-state population $\rho_\text{E}^\infty$ at several P$_0$ in Fig.~\ref{fig:onoff}(h) suggests the optimal resonant power for the maximum RF given a gating power. 


Now we briefly discuss the photophysics of the GeV system by comparing it to NV centers in diamond \cite{fu_conversion_2010, waldherr_dark_2011} and InGaAs self-assembled quantum dots (QD) \cite{nguyen_optically_2012, chen_characterization_2016}, where a similar phenomenon has been observed. For both systems, the dark state has been identified as a differently charged species of the emitter, specifically, positively charged QD \cite{nguyen_photoneutralization_2013} and neutrally charged NV center \cite{aslam_photo-induced_2013}. It is hence plausible that the dark state of the GeV center is also a differently charged state (i.e., neutral) \cite{thiering_ab_2018}.
For all three systems, the gating of RF can be achieved by employing a small amount of non-resonant beam. The mechanism for NV centers and QDs involves a local free-charge-carrier bath produced by the light, 
which can modify the charge dynamics of the emitter in favor of resonant excitation. We argue a similar mechanism for GeV center as long as non-resonant laser is employed. The linear power dependence of $k_\text{DG}$ and $k_\text{GD}$ [Fig.~\ref{fig:onoff}(d)] also supports this argument. 
On the other hand, the shelving mechanism induced by resonant pumping is different. For QDs, no such a shelving channel is reported. For NV centers, a two-photon process is involved based on the quadratic power dependence of the dynamical rates \cite{waldherr_dark_2011, siyushev_optically_2013}. For GeV center, a two-step shelving mechanism pivot by a meta-stable state and non-radiative decay channel is identified in this Letter. 
Finally, the decrease of rates $k_\text{GD}$ and $k_\text{DG}$ in Fig.~\ref{fig:onoff}(g) is possibly related to the decrease of free charge carrier density, caused by the presence of more charge traps in the area as induced by a stronger resonant beam \cite{GeV_supp_2019}. 


In summary, we demonstrated the shelving effect induced by the resonant laser in GeV centers, which can be counteracted by introducing a weak non-resonant repumping laser. The dynamics of shelving and gating can be quantitatively explained by the presence of a dark state, while the identity of this dark state warrants future investigation. We stress that this gating phenomenon is quite general and ubiquitous, not limited to the center investigated in this Letter \cite{GeV_supp_2019}. The recovery and stabilization of the RF could be useful for quantum information science and scalable quantum photonics, such as spin-photon entanglement \cite{togan_quantum_2010, de_greve_quantum-dot_2012} and photon photon interferences \cite{sipahigil_indistinguishable_2014}.

\begin{acknowledgments}
We acknowledge Singapore NRF fellowship grant (NRF-NRFF2015-03) and its Competitive Research Program (CRP Award No. NRF-CRP14-2014-02), Singapore Ministry of Education (MOE2016-T2-2-077, MOE2016-T2-1-163 and MOE2016-T3-1-006 (S)),  A*Star  QTE  programme  and  a  NTU start-up  grant  (M4081441), the Australian Research council (via DP180100077), the Asian Office of Aerospace Research and Development grant FA2386-17-1-4064, the Office of Naval Research Global(N62909-18-1-2025) and the AFAiiR node of the NCRIS Heavy Ion Capability for access to ion-implantation/ion-beam analysis facilities.

\end{acknowledgments}

\bibliography{./GeV_blinking_library_v6}

\begin{thebibliography}{36}%
\makeatletter
\providecommand \@ifxundefined [1]{%
 \@ifx{#1\undefined}
}%
\providecommand \@ifnum [1]{%
 \ifnum #1\expandafter \@firstoftwo
 \else \expandafter \@secondoftwo
 \fi
}%
\providecommand \@ifx [1]{%
 \ifx #1\expandafter \@firstoftwo
 \else \expandafter \@secondoftwo
 \fi
}%
\providecommand \natexlab [1]{#1}%
\providecommand \enquote  [1]{``#1''}%
\providecommand \bibnamefont  [1]{#1}%
\providecommand \bibfnamefont [1]{#1}%
\providecommand \citenamefont [1]{#1}%
\providecommand \href@noop [0]{\@secondoftwo}%
\providecommand \href [0]{\begingroup \@sanitize@url \@href}%
\providecommand \@href[1]{\@@startlink{#1}\@@href}%
\providecommand \@@href[1]{\endgroup#1\@@endlink}%
\providecommand \@sanitize@url [0]{\catcode `\\12\catcode `\$12\catcode
  `\&12\catcode `\#12\catcode `\^12\catcode `\_12\catcode `\%12\relax}%
\providecommand \@@startlink[1]{}%
\providecommand \@@endlink[0]{}%
\providecommand \url  [0]{\begingroup\@sanitize@url \@url }%
\providecommand \@url [1]{\endgroup\@href {#1}{\urlprefix }}%
\providecommand \urlprefix  [0]{URL }%
\providecommand \Eprint [0]{\href }%
\providecommand \doibase [0]{http://dx.doi.org/}%
\providecommand \selectlanguage [0]{\@gobble}%
\providecommand \bibinfo  [0]{\@secondoftwo}%
\providecommand \bibfield  [0]{\@secondoftwo}%
\providecommand \translation [1]{[#1]}%
\providecommand \BibitemOpen [0]{}%
\providecommand \bibitemStop [0]{}%
\providecommand \bibitemNoStop [0]{.\EOS\space}%
\providecommand \EOS [0]{\spacefactor3000\relax}%
\providecommand \BibitemShut  [1]{\csname bibitem#1\endcsname}%
\let\auto@bib@innerbib\@empty
\bibitem [{\citenamefont {Aharonovich}\ and\ \citenamefont
  {Neu}(2014)}]{aharonovich_diamond_2014}%
  \BibitemOpen
  \bibfield  {author} {\bibinfo {author} {\bibfnamefont {I.}~\bibnamefont
  {Aharonovich}}\ and\ \bibinfo {author} {\bibfnamefont {E.}~\bibnamefont
  {Neu}},\ }\href {\doibase 10.1002/adom.201400189} {\bibfield  {journal}
  {\bibinfo  {journal} {Advanced Optical Materials}\ }\textbf {\bibinfo
  {volume} {2}},\ \bibinfo {pages} {911} (\bibinfo {year} {2014})}\BibitemShut
  {NoStop}%
\bibitem [{\citenamefont {Atatüre}\ \emph {et~al.}(2018)\citenamefont
  {Atatüre}, \citenamefont {Englund}, \citenamefont {Vamivakas}, \citenamefont
  {Lee},\ and\ \citenamefont {Wrachtrup}}]{atature_material_2018}%
  \BibitemOpen
  \bibfield  {author} {\bibinfo {author} {\bibfnamefont {M.}~\bibnamefont
  {Atatüre}}, \bibinfo {author} {\bibfnamefont {D.}~\bibnamefont {Englund}},
  \bibinfo {author} {\bibfnamefont {N.}~\bibnamefont {Vamivakas}}, \bibinfo
  {author} {\bibfnamefont {S.-Y.}\ \bibnamefont {Lee}}, \ and\ \bibinfo
  {author} {\bibfnamefont {J.}~\bibnamefont {Wrachtrup}},\ }\href {\doibase
  10.1038/s41578-018-0008-9} {\bibfield  {journal} {\bibinfo  {journal} {Nature
  Reviews Materials}\ }\textbf {\bibinfo {volume} {3}},\ \bibinfo {pages} {38}
  (\bibinfo {year} {2018})}\BibitemShut {NoStop}%
\bibitem [{\citenamefont {Neu}\ \emph {et~al.}(2011{\natexlab{a}})\citenamefont
  {Neu}, \citenamefont {Steinmetz}, \citenamefont {Riedrich-Möller},
  \citenamefont {Gsell}, \citenamefont {Fischer}, \citenamefont {{Matthias
  Schreck}},\ and\ \citenamefont {Becher}}]{neu_single_2011}%
  \BibitemOpen
  \bibfield  {author} {\bibinfo {author} {\bibfnamefont {E.}~\bibnamefont
  {Neu}}, \bibinfo {author} {\bibfnamefont {D.}~\bibnamefont {Steinmetz}},
  \bibinfo {author} {\bibfnamefont {J.}~\bibnamefont {Riedrich-Möller}},
  \bibinfo {author} {\bibfnamefont {S.}~\bibnamefont {Gsell}}, \bibinfo
  {author} {\bibfnamefont {M.}~\bibnamefont {Fischer}}, \bibinfo {author}
  {\bibnamefont {{Matthias Schreck}}}, \ and\ \bibinfo {author} {\bibfnamefont
  {C.}~\bibnamefont {Becher}},\ }\href {\doibase 10.1088/1367-2630/13/2/025012}
  {\bibfield  {journal} {\bibinfo  {journal} {New Journal of Physics}\ }\textbf
  {\bibinfo {volume} {13}},\ \bibinfo {pages} {025012} (\bibinfo {year}
  {2011}{\natexlab{a}})}\BibitemShut {NoStop}%
\bibitem [{\citenamefont {Iwasaki}\ \emph {et~al.}(2015)\citenamefont
  {Iwasaki}, \citenamefont {Ishibashi}, \citenamefont {Miyamoto}, \citenamefont
  {Doi}, \citenamefont {Kobayashi}, \citenamefont {Miyazaki}, \citenamefont
  {Tahara}, \citenamefont {Jahnke}, \citenamefont {Rogers}, \citenamefont
  {Naydenov}, \citenamefont {Jelezko}, \citenamefont {Yamasaki}, \citenamefont
  {Nagamachi}, \citenamefont {Inubushi}, \citenamefont {Mizuochi},\ and\
  \citenamefont {Hatano}}]{iwasaki_germanium-vacancy_2015}%
  \BibitemOpen
  \bibfield  {author} {\bibinfo {author} {\bibfnamefont {T.}~\bibnamefont
  {Iwasaki}}, \bibinfo {author} {\bibfnamefont {F.}~\bibnamefont {Ishibashi}},
  \bibinfo {author} {\bibfnamefont {Y.}~\bibnamefont {Miyamoto}}, \bibinfo
  {author} {\bibfnamefont {Y.}~\bibnamefont {Doi}}, \bibinfo {author}
  {\bibfnamefont {S.}~\bibnamefont {Kobayashi}}, \bibinfo {author}
  {\bibfnamefont {T.}~\bibnamefont {Miyazaki}}, \bibinfo {author}
  {\bibfnamefont {K.}~\bibnamefont {Tahara}}, \bibinfo {author} {\bibfnamefont
  {K.~D.}\ \bibnamefont {Jahnke}}, \bibinfo {author} {\bibfnamefont {L.~J.}\
  \bibnamefont {Rogers}}, \bibinfo {author} {\bibfnamefont {B.}~\bibnamefont
  {Naydenov}}, \bibinfo {author} {\bibfnamefont {F.}~\bibnamefont {Jelezko}},
  \bibinfo {author} {\bibfnamefont {S.}~\bibnamefont {Yamasaki}}, \bibinfo
  {author} {\bibfnamefont {S.}~\bibnamefont {Nagamachi}}, \bibinfo {author}
  {\bibfnamefont {T.}~\bibnamefont {Inubushi}}, \bibinfo {author}
  {\bibfnamefont {N.}~\bibnamefont {Mizuochi}}, \ and\ \bibinfo {author}
  {\bibfnamefont {M.}~\bibnamefont {Hatano}},\ }\href {\doibase
  10.1038/srep12882} {\bibfield  {journal} {\bibinfo  {journal} {Scientific
  Reports}\ }\textbf {\bibinfo {volume} {5}},\ \bibinfo {pages} {12882}
  (\bibinfo {year} {2015})}\BibitemShut {NoStop}%
\bibitem [{\citenamefont {Palyanov}\ \emph {et~al.}(2015)\citenamefont
  {Palyanov}, \citenamefont {Kupriyanov}, \citenamefont {Borzdov},\ and\
  \citenamefont {Surovtsev}}]{palyanov_germanium:_2015}%
  \BibitemOpen
  \bibfield  {author} {\bibinfo {author} {\bibfnamefont {Y.~N.}\ \bibnamefont
  {Palyanov}}, \bibinfo {author} {\bibfnamefont {I.~N.}\ \bibnamefont
  {Kupriyanov}}, \bibinfo {author} {\bibfnamefont {Y.~M.}\ \bibnamefont
  {Borzdov}}, \ and\ \bibinfo {author} {\bibfnamefont {N.~V.}\ \bibnamefont
  {Surovtsev}},\ }\href {\doibase 10.1038/srep14789} {\bibfield  {journal}
  {\bibinfo  {journal} {Scientific Reports}\ }\textbf {\bibinfo {volume} {5}},\
  \bibinfo {pages} {14789} (\bibinfo {year} {2015})}\BibitemShut {NoStop}%
\bibitem [{\citenamefont {Häußler}\ \emph {et~al.}(2017)\citenamefont
  {Häußler}, \citenamefont {Thiering}, \citenamefont {Dietrich},
  \citenamefont {Waasem}, \citenamefont {Teraji}, \citenamefont {Isoya},
  \citenamefont {{Takayuki Iwasaki}}, \citenamefont {Hatano}, \citenamefont
  {Jelezko}, \citenamefont {Gali},\ and\ \citenamefont
  {Kubanek}}]{hausler_photoluminescence_2017}%
  \BibitemOpen
  \bibfield  {author} {\bibinfo {author} {\bibfnamefont {S.}~\bibnamefont
  {Häußler}}, \bibinfo {author} {\bibfnamefont {G.}~\bibnamefont {Thiering}},
  \bibinfo {author} {\bibfnamefont {A.}~\bibnamefont {Dietrich}}, \bibinfo
  {author} {\bibfnamefont {N.}~\bibnamefont {Waasem}}, \bibinfo {author}
  {\bibfnamefont {T.}~\bibnamefont {Teraji}}, \bibinfo {author} {\bibfnamefont
  {J.}~\bibnamefont {Isoya}}, \bibinfo {author} {\bibnamefont {{Takayuki
  Iwasaki}}}, \bibinfo {author} {\bibfnamefont {M.}~\bibnamefont {Hatano}},
  \bibinfo {author} {\bibfnamefont {F.}~\bibnamefont {Jelezko}}, \bibinfo
  {author} {\bibfnamefont {A.}~\bibnamefont {Gali}}, \ and\ \bibinfo {author}
  {\bibfnamefont {A.}~\bibnamefont {Kubanek}},\ }\href {\doibase
  10.1088/1367-2630/aa73e5} {\bibfield  {journal} {\bibinfo  {journal} {New
  Journal of Physics}\ }\textbf {\bibinfo {volume} {19}},\ \bibinfo {pages}
  {063036} (\bibinfo {year} {2017})}\BibitemShut {NoStop}%
\bibitem [{\citenamefont {Bhaskar}\ \emph {et~al.}(2017)\citenamefont
  {Bhaskar}, \citenamefont {Sukachev}, \citenamefont {Sipahigil}, \citenamefont
  {Evans}, \citenamefont {Burek}, \citenamefont {Nguyen}, \citenamefont
  {Rogers}, \citenamefont {Siyushev}, \citenamefont {Metsch}, \citenamefont
  {Park}, \citenamefont {Jelezko}, \citenamefont {Lončar},\ and\ \citenamefont
  {Lukin}}]{bhaskar_quantum_2017}%
  \BibitemOpen
  \bibfield  {author} {\bibinfo {author} {\bibfnamefont {M.}~\bibnamefont
  {Bhaskar}}, \bibinfo {author} {\bibfnamefont {D.}~\bibnamefont {Sukachev}},
  \bibinfo {author} {\bibfnamefont {A.}~\bibnamefont {Sipahigil}}, \bibinfo
  {author} {\bibfnamefont {R.}~\bibnamefont {Evans}}, \bibinfo {author}
  {\bibfnamefont {M.}~\bibnamefont {Burek}}, \bibinfo {author} {\bibfnamefont
  {C.}~\bibnamefont {Nguyen}}, \bibinfo {author} {\bibfnamefont
  {L.}~\bibnamefont {Rogers}}, \bibinfo {author} {\bibfnamefont
  {P.}~\bibnamefont {Siyushev}}, \bibinfo {author} {\bibfnamefont
  {M.}~\bibnamefont {Metsch}}, \bibinfo {author} {\bibfnamefont
  {H.}~\bibnamefont {Park}}, \bibinfo {author} {\bibfnamefont {F.}~\bibnamefont
  {Jelezko}}, \bibinfo {author} {\bibfnamefont {M.}~\bibnamefont {Lončar}}, \
  and\ \bibinfo {author} {\bibfnamefont {M.}~\bibnamefont {Lukin}},\ }\href
  {\doibase 10.1103/PhysRevLett.118.223603} {\bibfield  {journal} {\bibinfo
  {journal} {Physical Review Letters}\ }\textbf {\bibinfo {volume} {118}},\
  \bibinfo {pages} {223603} (\bibinfo {year} {2017})}\BibitemShut {NoStop}%
\bibitem [{\citenamefont {Siampour}\ \emph {et~al.}(2018)\citenamefont
  {Siampour}, \citenamefont {Kumar}, \citenamefont {Davydov}, \citenamefont
  {Kulikova}, \citenamefont {Agafonov},\ and\ \citenamefont
  {Bozhevolnyi}}]{siampour_-chip_2018}%
  \BibitemOpen
  \bibfield  {author} {\bibinfo {author} {\bibfnamefont {H.}~\bibnamefont
  {Siampour}}, \bibinfo {author} {\bibfnamefont {S.}~\bibnamefont {Kumar}},
  \bibinfo {author} {\bibfnamefont {V.~A.}\ \bibnamefont {Davydov}}, \bibinfo
  {author} {\bibfnamefont {L.~F.}\ \bibnamefont {Kulikova}}, \bibinfo {author}
  {\bibfnamefont {V.~N.}\ \bibnamefont {Agafonov}}, \ and\ \bibinfo {author}
  {\bibfnamefont {S.~I.}\ \bibnamefont {Bozhevolnyi}},\ }\href {\doibase
  10.1038/s41377-018-0062-5} {\bibfield  {journal} {\bibinfo  {journal} {Light:
  Science \& Applications}\ }\textbf {\bibinfo {volume} {7}},\ \bibinfo {pages}
  {61} (\bibinfo {year} {2018})}\BibitemShut {NoStop}%
\bibitem [{\citenamefont {Bray}\ \emph {et~al.}(2018)\citenamefont {Bray},
  \citenamefont {Regan}, \citenamefont {Trycz}, \citenamefont {Previdi},
  \citenamefont {Seniutinas}, \citenamefont {Ganesan}, \citenamefont
  {Kianinia}, \citenamefont {Kim},\ and\ \citenamefont
  {Aharonovich}}]{bray_single_2018}%
  \BibitemOpen
  \bibfield  {author} {\bibinfo {author} {\bibfnamefont {K.}~\bibnamefont
  {Bray}}, \bibinfo {author} {\bibfnamefont {B.}~\bibnamefont {Regan}},
  \bibinfo {author} {\bibfnamefont {A.}~\bibnamefont {Trycz}}, \bibinfo
  {author} {\bibfnamefont {R.}~\bibnamefont {Previdi}}, \bibinfo {author}
  {\bibfnamefont {G.}~\bibnamefont {Seniutinas}}, \bibinfo {author}
  {\bibfnamefont {K.}~\bibnamefont {Ganesan}}, \bibinfo {author} {\bibfnamefont
  {M.}~\bibnamefont {Kianinia}}, \bibinfo {author} {\bibfnamefont
  {S.}~\bibnamefont {Kim}}, \ and\ \bibinfo {author} {\bibfnamefont
  {I.}~\bibnamefont {Aharonovich}},\ }\href {\doibase
  10.1021/acsphotonics.8b00930} {\bibfield  {journal} {\bibinfo  {journal} {ACS
  Photonics}\ }\textbf {\bibinfo {volume} {5}},\ \bibinfo {pages} {4817}
  (\bibinfo {year} {2018})}\BibitemShut {NoStop}%
\bibitem [{\citenamefont {Hepp}\ \emph {et~al.}(2014)\citenamefont {Hepp},
  \citenamefont {Müller}, \citenamefont {Waselowski}, \citenamefont {Becker},
  \citenamefont {Pingault}, \citenamefont {Sternschulte}, \citenamefont
  {Steinmüller-Nethl}, \citenamefont {Gali}, \citenamefont {Maze},
  \citenamefont {Atatüre},\ and\ \citenamefont
  {Becher}}]{hepp_electronic_2014}%
  \BibitemOpen
  \bibfield  {author} {\bibinfo {author} {\bibfnamefont {C.}~\bibnamefont
  {Hepp}}, \bibinfo {author} {\bibfnamefont {T.}~\bibnamefont {Müller}},
  \bibinfo {author} {\bibfnamefont {V.}~\bibnamefont {Waselowski}}, \bibinfo
  {author} {\bibfnamefont {J.~N.}\ \bibnamefont {Becker}}, \bibinfo {author}
  {\bibfnamefont {B.}~\bibnamefont {Pingault}}, \bibinfo {author}
  {\bibfnamefont {H.}~\bibnamefont {Sternschulte}}, \bibinfo {author}
  {\bibfnamefont {D.}~\bibnamefont {Steinmüller-Nethl}}, \bibinfo {author}
  {\bibfnamefont {A.}~\bibnamefont {Gali}}, \bibinfo {author} {\bibfnamefont
  {J.~R.}\ \bibnamefont {Maze}}, \bibinfo {author} {\bibfnamefont
  {M.}~\bibnamefont {Atatüre}}, \ and\ \bibinfo {author} {\bibfnamefont
  {C.}~\bibnamefont {Becher}},\ }\href {\doibase
  10.1103/PhysRevLett.112.036405} {\bibfield  {journal} {\bibinfo  {journal}
  {Physical Review Letters}\ }\textbf {\bibinfo {volume} {112}},\ \bibinfo
  {pages} {036405} (\bibinfo {year} {2014})}\BibitemShut {NoStop}%
\bibitem [{\citenamefont {Sipahigil}\ \emph {et~al.}(2014)\citenamefont
  {Sipahigil}, \citenamefont {Jahnke}, \citenamefont {Rogers}, \citenamefont
  {Teraji}, \citenamefont {Isoya}, \citenamefont {Zibrov}, \citenamefont
  {Jelezko},\ and\ \citenamefont {Lukin}}]{sipahigil_indistinguishable_2014}%
  \BibitemOpen
  \bibfield  {author} {\bibinfo {author} {\bibfnamefont {A.}~\bibnamefont
  {Sipahigil}}, \bibinfo {author} {\bibfnamefont {K.}~\bibnamefont {Jahnke}},
  \bibinfo {author} {\bibfnamefont {L.}~\bibnamefont {Rogers}}, \bibinfo
  {author} {\bibfnamefont {T.}~\bibnamefont {Teraji}}, \bibinfo {author}
  {\bibfnamefont {J.}~\bibnamefont {Isoya}}, \bibinfo {author} {\bibfnamefont
  {A.}~\bibnamefont {Zibrov}}, \bibinfo {author} {\bibfnamefont
  {F.}~\bibnamefont {Jelezko}}, \ and\ \bibinfo {author} {\bibfnamefont
  {M.}~\bibnamefont {Lukin}},\ }\href {\doibase 10.1103/PhysRevLett.113.113602}
  {\bibfield  {journal} {\bibinfo  {journal} {Physical Review Letters}\
  }\textbf {\bibinfo {volume} {113}},\ \bibinfo {pages} {113602} (\bibinfo
  {year} {2014})}\BibitemShut {NoStop}%
\bibitem [{\citenamefont {Neu}\ \emph {et~al.}(2011{\natexlab{b}})\citenamefont
  {Neu}, \citenamefont {Fischer}, \citenamefont {Gsell}, \citenamefont
  {Schreck},\ and\ \citenamefont {Becher}}]{neu_fluorescence_2011}%
  \BibitemOpen
  \bibfield  {author} {\bibinfo {author} {\bibfnamefont {E.}~\bibnamefont
  {Neu}}, \bibinfo {author} {\bibfnamefont {M.}~\bibnamefont {Fischer}},
  \bibinfo {author} {\bibfnamefont {S.}~\bibnamefont {Gsell}}, \bibinfo
  {author} {\bibfnamefont {M.}~\bibnamefont {Schreck}}, \ and\ \bibinfo
  {author} {\bibfnamefont {C.}~\bibnamefont {Becher}},\ }\href {\doibase
  10.1103/PhysRevB.84.205211} {\bibfield  {journal} {\bibinfo  {journal}
  {Physical Review B}\ }\textbf {\bibinfo {volume} {84}},\ \bibinfo {pages}
  {205211} (\bibinfo {year} {2011}{\natexlab{b}})}\BibitemShut {NoStop}%
\bibitem [{\citenamefont {Duan}\ \emph {et~al.}(2001)\citenamefont {Duan},
  \citenamefont {Lukin}, \citenamefont {Cirac},\ and\ \citenamefont
  {Zoller}}]{duan_long-distance_2001}%
  \BibitemOpen
  \bibfield  {author} {\bibinfo {author} {\bibfnamefont {L.-M.}\ \bibnamefont
  {Duan}}, \bibinfo {author} {\bibfnamefont {M.~D.}\ \bibnamefont {Lukin}},
  \bibinfo {author} {\bibfnamefont {J.~I.}\ \bibnamefont {Cirac}}, \ and\
  \bibinfo {author} {\bibfnamefont {P.}~\bibnamefont {Zoller}},\ }\href
  {\doibase 10.1038/35106500} {\bibfield  {journal} {\bibinfo  {journal}
  {Nature}\ }\textbf {\bibinfo {volume} {414}},\ \bibinfo {pages} {413}
  (\bibinfo {year} {2001})}\BibitemShut {NoStop}%
\bibitem [{\citenamefont {Bouwmeester}\ \emph {et~al.}(1997)\citenamefont
  {Bouwmeester}, \citenamefont {Pan}, \citenamefont {Mattle}, \citenamefont
  {Eibl}, \citenamefont {Weinfurter},\ and\ \citenamefont
  {Zeilinger}}]{bouwmeester_experimental_1997}%
  \BibitemOpen
  \bibfield  {author} {\bibinfo {author} {\bibfnamefont {D.}~\bibnamefont
  {Bouwmeester}}, \bibinfo {author} {\bibfnamefont {J.-W.}\ \bibnamefont
  {Pan}}, \bibinfo {author} {\bibfnamefont {K.}~\bibnamefont {Mattle}},
  \bibinfo {author} {\bibfnamefont {M.}~\bibnamefont {Eibl}}, \bibinfo {author}
  {\bibfnamefont {H.}~\bibnamefont {Weinfurter}}, \ and\ \bibinfo {author}
  {\bibfnamefont {A.}~\bibnamefont {Zeilinger}},\ }\href {\doibase
  10.1038/37539} {\bibfield  {journal} {\bibinfo  {journal} {Nature}\ }\textbf
  {\bibinfo {volume} {390}},\ \bibinfo {pages} {575} (\bibinfo {year}
  {1997})}\BibitemShut {NoStop}%
\bibitem [{\citenamefont {Pan}\ \emph {et~al.}(1998)\citenamefont {Pan},
  \citenamefont {Bouwmeester}, \citenamefont {Weinfurter},\ and\ \citenamefont
  {Zeilinger}}]{pan_experimental_1998}%
  \BibitemOpen
  \bibfield  {author} {\bibinfo {author} {\bibfnamefont {J.-W.}\ \bibnamefont
  {Pan}}, \bibinfo {author} {\bibfnamefont {D.}~\bibnamefont {Bouwmeester}},
  \bibinfo {author} {\bibfnamefont {H.}~\bibnamefont {Weinfurter}}, \ and\
  \bibinfo {author} {\bibfnamefont {A.}~\bibnamefont {Zeilinger}},\ }\href
  {\doibase 10.1103/PhysRevLett.80.3891} {\bibfield  {journal} {\bibinfo
  {journal} {Physical Review Letters}\ }\textbf {\bibinfo {volume} {80}},\
  \bibinfo {pages} {3891} (\bibinfo {year} {1998})}\BibitemShut {NoStop}%
\bibitem [{\citenamefont {Doherty}\ \emph {et~al.}(2013)\citenamefont
  {Doherty}, \citenamefont {Manson}, \citenamefont {Delaney}, \citenamefont
  {Jelezko}, \citenamefont {Wrachtrup},\ and\ \citenamefont
  {Hollenberg}}]{doherty_nitrogen-vacancy_2013}%
  \BibitemOpen
  \bibfield  {author} {\bibinfo {author} {\bibfnamefont {M.~W.}\ \bibnamefont
  {Doherty}}, \bibinfo {author} {\bibfnamefont {N.~B.}\ \bibnamefont {Manson}},
  \bibinfo {author} {\bibfnamefont {P.}~\bibnamefont {Delaney}}, \bibinfo
  {author} {\bibfnamefont {F.}~\bibnamefont {Jelezko}}, \bibinfo {author}
  {\bibfnamefont {J.}~\bibnamefont {Wrachtrup}}, \ and\ \bibinfo {author}
  {\bibfnamefont {L.~C.~L.}\ \bibnamefont {Hollenberg}},\ }\href {\doibase
  10.1016/j.physrep.2013.02.001} {\bibfield  {journal} {\bibinfo  {journal}
  {Physics Reports}\ }\bibinfo {series} {The nitrogen-vacancy colour centre in
  diamond},\ \textbf {\bibinfo {volume} {528}},\ \bibinfo {pages} {1} (\bibinfo
  {year} {2013})}\BibitemShut {NoStop}%
\bibitem [{\citenamefont {Waldherr}\ \emph {et~al.}(2011)\citenamefont
  {Waldherr}, \citenamefont {Beck}, \citenamefont {Steiner}, \citenamefont
  {Neumann}, \citenamefont {Gali}, \citenamefont {Frauenheim}, \citenamefont
  {Jelezko},\ and\ \citenamefont {Wrachtrup}}]{waldherr_dark_2011}%
  \BibitemOpen
  \bibfield  {author} {\bibinfo {author} {\bibfnamefont {G.}~\bibnamefont
  {Waldherr}}, \bibinfo {author} {\bibfnamefont {J.}~\bibnamefont {Beck}},
  \bibinfo {author} {\bibfnamefont {M.}~\bibnamefont {Steiner}}, \bibinfo
  {author} {\bibfnamefont {P.}~\bibnamefont {Neumann}}, \bibinfo {author}
  {\bibfnamefont {A.}~\bibnamefont {Gali}}, \bibinfo {author} {\bibfnamefont
  {T.}~\bibnamefont {Frauenheim}}, \bibinfo {author} {\bibfnamefont
  {F.}~\bibnamefont {Jelezko}}, \ and\ \bibinfo {author} {\bibfnamefont
  {J.}~\bibnamefont {Wrachtrup}},\ }\href {\doibase
  10.1103/PhysRevLett.106.157601} {\bibfield  {journal} {\bibinfo  {journal}
  {Physical Review Letters}\ }\textbf {\bibinfo {volume} {106}},\ \bibinfo
  {pages} {157601} (\bibinfo {year} {2011})}\BibitemShut {NoStop}%
\bibitem [{\citenamefont {Siyushev}\ \emph {et~al.}(2013)\citenamefont
  {Siyushev}, \citenamefont {Pinto}, \citenamefont {Vörös}, \citenamefont
  {Gali}, \citenamefont {Jelezko},\ and\ \citenamefont
  {Wrachtrup}}]{siyushev_optically_2013}%
  \BibitemOpen
  \bibfield  {author} {\bibinfo {author} {\bibfnamefont {P.}~\bibnamefont
  {Siyushev}}, \bibinfo {author} {\bibfnamefont {H.}~\bibnamefont {Pinto}},
  \bibinfo {author} {\bibfnamefont {M.}~\bibnamefont {Vörös}}, \bibinfo
  {author} {\bibfnamefont {A.}~\bibnamefont {Gali}}, \bibinfo {author}
  {\bibfnamefont {F.}~\bibnamefont {Jelezko}}, \ and\ \bibinfo {author}
  {\bibfnamefont {J.}~\bibnamefont {Wrachtrup}},\ }\href {\doibase
  10.1103/PhysRevLett.110.167402} {\bibfield  {journal} {\bibinfo  {journal}
  {Physical Review Letters}\ }\textbf {\bibinfo {volume} {110}},\ \bibinfo
  {pages} {167402} (\bibinfo {year} {2013})}\BibitemShut {NoStop}%
\bibitem [{\citenamefont {Sukachev}\ \emph {et~al.}(2017)\citenamefont
  {Sukachev}, \citenamefont {Sipahigil}, \citenamefont {Nguyen}, \citenamefont
  {Bhaskar}, \citenamefont {Evans}, \citenamefont {Jelezko},\ and\
  \citenamefont {Lukin}}]{sukachev_silicon-vacancy_2017}%
  \BibitemOpen
  \bibfield  {author} {\bibinfo {author} {\bibfnamefont {D.}~\bibnamefont
  {Sukachev}}, \bibinfo {author} {\bibfnamefont {A.}~\bibnamefont {Sipahigil}},
  \bibinfo {author} {\bibfnamefont {C.}~\bibnamefont {Nguyen}}, \bibinfo
  {author} {\bibfnamefont {M.}~\bibnamefont {Bhaskar}}, \bibinfo {author}
  {\bibfnamefont {R.}~\bibnamefont {Evans}}, \bibinfo {author} {\bibfnamefont
  {F.}~\bibnamefont {Jelezko}}, \ and\ \bibinfo {author} {\bibfnamefont
  {M.}~\bibnamefont {Lukin}},\ }\href {\doibase 10.1103/PhysRevLett.119.223602}
  {\bibfield  {journal} {\bibinfo  {journal} {Physical Review Letters}\
  }\textbf {\bibinfo {volume} {119}},\ \bibinfo {pages} {223602} (\bibinfo
  {year} {2017})}\BibitemShut {NoStop}%
\bibitem [{\citenamefont {Vamivakas}\ \emph {et~al.}(2010)\citenamefont
  {Vamivakas}, \citenamefont {Lu}, \citenamefont {Matthiesen}, \citenamefont
  {Zhao}, \citenamefont {Fält}, \citenamefont {Badolato},\ and\ \citenamefont
  {Atatüre}}]{vamivakas_observation_2010}%
  \BibitemOpen
  \bibfield  {author} {\bibinfo {author} {\bibfnamefont {A.~N.}\ \bibnamefont
  {Vamivakas}}, \bibinfo {author} {\bibfnamefont {C.-Y.}\ \bibnamefont {Lu}},
  \bibinfo {author} {\bibfnamefont {C.}~\bibnamefont {Matthiesen}}, \bibinfo
  {author} {\bibfnamefont {Y.}~\bibnamefont {Zhao}}, \bibinfo {author}
  {\bibfnamefont {S.}~\bibnamefont {Fält}}, \bibinfo {author} {\bibfnamefont
  {A.}~\bibnamefont {Badolato}}, \ and\ \bibinfo {author} {\bibfnamefont
  {M.}~\bibnamefont {Atatüre}},\ }\href {\doibase 10.1038/nature09359}
  {\bibfield  {journal} {\bibinfo  {journal} {Nature}\ }\textbf {\bibinfo
  {volume} {467}},\ \bibinfo {pages} {297} (\bibinfo {year}
  {2010})}\BibitemShut {NoStop}%
\bibitem [{\citenamefont {Kimble}(2008)}]{kimble_quantum_2008}%
  \BibitemOpen
  \bibfield  {author} {\bibinfo {author} {\bibfnamefont {H.~J.}\ \bibnamefont
  {Kimble}},\ }\href {\doibase 10.1038/nature07127} {\bibfield  {journal}
  {\bibinfo  {journal} {Nature}\ }\textbf {\bibinfo {volume} {453}},\ \bibinfo
  {pages} {1023} (\bibinfo {year} {2008})}\BibitemShut {NoStop}%
\bibitem [{GeV()}]{GeV_supp_2019}%
  \BibitemOpen
  \href@noop {} {\bibinfo  {journal} {See Supplementary Material, which
  includes Refs. ...}\ }\BibitemShut {NoStop}%
\bibitem [{\citenamefont {Thiering}\ and\ \citenamefont
  {Gali}(2018)}]{thiering_ab_2018}%
  \BibitemOpen
\bibfield  {journal} {  }\bibfield  {author} {\bibinfo {author} {\bibfnamefont
  {G.}~\bibnamefont {Thiering}}\ and\ \bibinfo {author} {\bibfnamefont
  {A.}~\bibnamefont {Gali}},\ }\href {\doibase 10.1103/PhysRevX.8.021063}
  {\bibfield  {journal} {\bibinfo  {journal} {Physical Review X}\ }\textbf
  {\bibinfo {volume} {8}},\ \bibinfo {pages} {021063} (\bibinfo {year}
  {2018})}\BibitemShut {NoStop}%
\bibitem [{\citenamefont {Marseglia}\ \emph {et~al.}(2011)\citenamefont
  {Marseglia}, \citenamefont {Hadden}, \citenamefont {Stanley-Clarke},
  \citenamefont {Harrison}, \citenamefont {Patton}, \citenamefont {Ho},
  \citenamefont {Naydenov}, \citenamefont {Jelezko}, \citenamefont {Meijer},
  \citenamefont {Dolan}, \citenamefont {Smith}, \citenamefont {Rarity},\ and\
  \citenamefont {O’Brien}}]{marseglia_nanofabricated_2011}%
  \BibitemOpen
  \bibfield  {author} {\bibinfo {author} {\bibfnamefont {L.}~\bibnamefont
  {Marseglia}}, \bibinfo {author} {\bibfnamefont {J.~P.}\ \bibnamefont
  {Hadden}}, \bibinfo {author} {\bibfnamefont {A.~C.}\ \bibnamefont
  {Stanley-Clarke}}, \bibinfo {author} {\bibfnamefont {J.~P.}\ \bibnamefont
  {Harrison}}, \bibinfo {author} {\bibfnamefont {B.}~\bibnamefont {Patton}},
  \bibinfo {author} {\bibfnamefont {Y.-L.~D.}\ \bibnamefont {Ho}}, \bibinfo
  {author} {\bibfnamefont {B.}~\bibnamefont {Naydenov}}, \bibinfo {author}
  {\bibfnamefont {F.}~\bibnamefont {Jelezko}}, \bibinfo {author} {\bibfnamefont
  {J.}~\bibnamefont {Meijer}}, \bibinfo {author} {\bibfnamefont {P.~R.}\
  \bibnamefont {Dolan}}, \bibinfo {author} {\bibfnamefont {J.~M.}\ \bibnamefont
  {Smith}}, \bibinfo {author} {\bibfnamefont {J.~G.}\ \bibnamefont {Rarity}}, \
  and\ \bibinfo {author} {\bibfnamefont {J.~L.}\ \bibnamefont {O’Brien}},\
  }\href {\doibase 10.1063/1.3573870} {\bibfield  {journal} {\bibinfo
  {journal} {Applied Physics Letters}\ }\textbf {\bibinfo {volume} {98}},\
  \bibinfo {pages} {133107} (\bibinfo {year} {2011})}\BibitemShut {NoStop}%
\bibitem [{\citenamefont {Siyushev}\ \emph {et~al.}(2017)\citenamefont
  {Siyushev}, \citenamefont {Metsch}, \citenamefont {Ijaz}, \citenamefont
  {Binder}, \citenamefont {Bhaskar}, \citenamefont {Sukachev}, \citenamefont
  {Sipahigil}, \citenamefont {Evans}, \citenamefont {Nguyen}, \citenamefont
  {Lukin}, \citenamefont {Hemmer}, \citenamefont {Palyanov}, \citenamefont
  {Kupriyanov}, \citenamefont {Borzdov}, \citenamefont {Rogers},\ and\
  \citenamefont {Jelezko}}]{siyushev_optical_2017}%
  \BibitemOpen
  \bibfield  {author} {\bibinfo {author} {\bibfnamefont {P.}~\bibnamefont
  {Siyushev}}, \bibinfo {author} {\bibfnamefont {M.~H.}\ \bibnamefont
  {Metsch}}, \bibinfo {author} {\bibfnamefont {A.}~\bibnamefont {Ijaz}},
  \bibinfo {author} {\bibfnamefont {J.~M.}\ \bibnamefont {Binder}}, \bibinfo
  {author} {\bibfnamefont {M.~K.}\ \bibnamefont {Bhaskar}}, \bibinfo {author}
  {\bibfnamefont {D.~D.}\ \bibnamefont {Sukachev}}, \bibinfo {author}
  {\bibfnamefont {A.}~\bibnamefont {Sipahigil}}, \bibinfo {author}
  {\bibfnamefont {R.~E.}\ \bibnamefont {Evans}}, \bibinfo {author}
  {\bibfnamefont {C.~T.}\ \bibnamefont {Nguyen}}, \bibinfo {author}
  {\bibfnamefont {M.~D.}\ \bibnamefont {Lukin}}, \bibinfo {author}
  {\bibfnamefont {P.~R.}\ \bibnamefont {Hemmer}}, \bibinfo {author}
  {\bibfnamefont {Y.~N.}\ \bibnamefont {Palyanov}}, \bibinfo {author}
  {\bibfnamefont {I.~N.}\ \bibnamefont {Kupriyanov}}, \bibinfo {author}
  {\bibfnamefont {Y.~M.}\ \bibnamefont {Borzdov}}, \bibinfo {author}
  {\bibfnamefont {L.~J.}\ \bibnamefont {Rogers}}, \ and\ \bibinfo {author}
  {\bibfnamefont {F.}~\bibnamefont {Jelezko}},\ }\href {\doibase
  10.1103/PhysRevB.96.081201} {\bibfield  {journal} {\bibinfo  {journal}
  {Physical Review B}\ }\textbf {\bibinfo {volume} {96}},\ \bibinfo {pages}
  {081201} (\bibinfo {year} {2017})}\BibitemShut {NoStop}%
\bibitem [{\citenamefont {Nguyen}\ \emph {et~al.}(2013)\citenamefont {Nguyen},
  \citenamefont {Sallen}, \citenamefont {Abbarchi}, \citenamefont {Ferreira},
  \citenamefont {Voisin}, \citenamefont {Roussignol}, \citenamefont
  {Cassabois},\ and\ \citenamefont
  {Diederichs}}]{nguyen_photoneutralization_2013}%
  \BibitemOpen
  \bibfield  {author} {\bibinfo {author} {\bibfnamefont {H.~S.}\ \bibnamefont
  {Nguyen}}, \bibinfo {author} {\bibfnamefont {G.}~\bibnamefont {Sallen}},
  \bibinfo {author} {\bibfnamefont {M.}~\bibnamefont {Abbarchi}}, \bibinfo
  {author} {\bibfnamefont {R.}~\bibnamefont {Ferreira}}, \bibinfo {author}
  {\bibfnamefont {C.}~\bibnamefont {Voisin}}, \bibinfo {author} {\bibfnamefont
  {P.}~\bibnamefont {Roussignol}}, \bibinfo {author} {\bibfnamefont
  {G.}~\bibnamefont {Cassabois}}, \ and\ \bibinfo {author} {\bibfnamefont
  {C.}~\bibnamefont {Diederichs}},\ }\href {\doibase
  10.1103/PhysRevB.87.115305} {\bibfield  {journal} {\bibinfo  {journal}
  {Physical Review B}\ }\textbf {\bibinfo {volume} {87}},\ \bibinfo {pages}
  {115305} (\bibinfo {year} {2013})}\BibitemShut {NoStop}%
\bibitem [{\citenamefont {Delteil}\ \emph {et~al.}(2014)\citenamefont
  {Delteil}, \citenamefont {Gao}, \citenamefont {Fallahi}, \citenamefont
  {Miguel-Sanchez},\ and\ \citenamefont
  {Imamoğlu}}]{delteil_observation_2014}%
  \BibitemOpen
  \bibfield  {author} {\bibinfo {author} {\bibfnamefont {A.}~\bibnamefont
  {Delteil}}, \bibinfo {author} {\bibfnamefont {W.-b.}\ \bibnamefont {Gao}},
  \bibinfo {author} {\bibfnamefont {P.}~\bibnamefont {Fallahi}}, \bibinfo
  {author} {\bibfnamefont {J.}~\bibnamefont {Miguel-Sanchez}}, \ and\ \bibinfo
  {author} {\bibfnamefont {A.}~\bibnamefont {Imamoğlu}},\ }\href {\doibase
  10.1103/PhysRevLett.112.116802} {\bibfield  {journal} {\bibinfo  {journal}
  {Physical Review Letters}\ }\textbf {\bibinfo {volume} {112}},\ \bibinfo
  {pages} {116802} (\bibinfo {year} {2014})}\BibitemShut {NoStop}%
\bibitem [{\citenamefont {Wolters}\ \emph {et~al.}(2013)\citenamefont
  {Wolters}, \citenamefont {Sadzak}, \citenamefont {Schell}, \citenamefont
  {Schröder},\ and\ \citenamefont {Benson}}]{wolters_measurement_2013}%
  \BibitemOpen
  \bibfield  {author} {\bibinfo {author} {\bibfnamefont {J.}~\bibnamefont
  {Wolters}}, \bibinfo {author} {\bibfnamefont {N.}~\bibnamefont {Sadzak}},
  \bibinfo {author} {\bibfnamefont {A.~W.}\ \bibnamefont {Schell}}, \bibinfo
  {author} {\bibfnamefont {T.}~\bibnamefont {Schröder}}, \ and\ \bibinfo
  {author} {\bibfnamefont {O.}~\bibnamefont {Benson}},\ }\href {\doibase
  10.1103/PhysRevLett.110.027401} {\bibfield  {journal} {\bibinfo  {journal}
  {Physical Review Letters}\ }\textbf {\bibinfo {volume} {110}},\ \bibinfo
  {pages} {027401} (\bibinfo {year} {2013})}\BibitemShut {NoStop}%
\bibitem [{\citenamefont {Becker}\ \emph {et~al.}(2018)\citenamefont {Becker},
  \citenamefont {Pingault}, \citenamefont {Groß}, \citenamefont {Gündoğan},
  \citenamefont {Kukharchyk}, \citenamefont {Markham}, \citenamefont {Edmonds},
  \citenamefont {Atatüre}, \citenamefont {Bushev},\ and\ \citenamefont
  {Becher}}]{becker_all-optical_2018}%
  \BibitemOpen
  \bibfield  {author} {\bibinfo {author} {\bibfnamefont {J.~N.}\ \bibnamefont
  {Becker}}, \bibinfo {author} {\bibfnamefont {B.}~\bibnamefont {Pingault}},
  \bibinfo {author} {\bibfnamefont {D.}~\bibnamefont {Groß}}, \bibinfo
  {author} {\bibfnamefont {M.}~\bibnamefont {Gündoğan}}, \bibinfo {author}
  {\bibfnamefont {N.}~\bibnamefont {Kukharchyk}}, \bibinfo {author}
  {\bibfnamefont {M.}~\bibnamefont {Markham}}, \bibinfo {author} {\bibfnamefont
  {A.}~\bibnamefont {Edmonds}}, \bibinfo {author} {\bibfnamefont
  {M.}~\bibnamefont {Atatüre}}, \bibinfo {author} {\bibfnamefont
  {P.}~\bibnamefont {Bushev}}, \ and\ \bibinfo {author} {\bibfnamefont
  {C.}~\bibnamefont {Becher}},\ }\href {\doibase
  10.1103/PhysRevLett.120.053603} {\bibfield  {journal} {\bibinfo  {journal}
  {Physical Review Letters}\ }\textbf {\bibinfo {volume} {120}},\ \bibinfo
  {pages} {053603} (\bibinfo {year} {2018})}\BibitemShut {NoStop}%
\bibitem [{\citenamefont {Boldyrev}\ \emph {et~al.}(2018)\citenamefont
  {Boldyrev}, \citenamefont {Mavrin}, \citenamefont {Sherin},\ and\
  \citenamefont {Popova}}]{boldyrev_bright_2018}%
  \BibitemOpen
  \bibfield  {author} {\bibinfo {author} {\bibfnamefont {K.~N.}\ \bibnamefont
  {Boldyrev}}, \bibinfo {author} {\bibfnamefont {B.~N.}\ \bibnamefont
  {Mavrin}}, \bibinfo {author} {\bibfnamefont {P.~S.}\ \bibnamefont {Sherin}},
  \ and\ \bibinfo {author} {\bibfnamefont {M.~N.}\ \bibnamefont {Popova}},\
  }\href {\doibase 10.1016/j.jlumin.2017.07.031} {\bibfield  {journal}
  {\bibinfo  {journal} {Journal of Luminescence}\ }\bibinfo {series} {Special
  {Issue} in honor of {Professor} {Renata} {Reisfeld} for her outstanding
  contributions to luminescent inorganic glasses},\ \textbf {\bibinfo {volume}
  {193}},\ \bibinfo {pages} {119} (\bibinfo {year} {2018})}\BibitemShut
  {NoStop}%
\bibitem [{\citenamefont {Fu}\ \emph {et~al.}(2010)\citenamefont {Fu},
  \citenamefont {Santori}, \citenamefont {Barclay},\ and\ \citenamefont
  {Beausoleil}}]{fu_conversion_2010}%
  \BibitemOpen
  \bibfield  {author} {\bibinfo {author} {\bibfnamefont {K.-M.~C.}\
  \bibnamefont {Fu}}, \bibinfo {author} {\bibfnamefont {C.}~\bibnamefont
  {Santori}}, \bibinfo {author} {\bibfnamefont {P.~E.}\ \bibnamefont
  {Barclay}}, \ and\ \bibinfo {author} {\bibfnamefont {R.~G.}\ \bibnamefont
  {Beausoleil}},\ }\href {\doibase 10.1063/1.3364135} {\bibfield  {journal}
  {\bibinfo  {journal} {Applied Physics Letters}\ }\textbf {\bibinfo {volume}
  {96}},\ \bibinfo {pages} {121907} (\bibinfo {year} {2010})}\BibitemShut
  {NoStop}%
\bibitem [{\citenamefont {Nguyen}\ \emph {et~al.}(2012)\citenamefont {Nguyen},
  \citenamefont {Sallen}, \citenamefont {Voisin}, \citenamefont {Roussignol},
  \citenamefont {Diederichs},\ and\ \citenamefont
  {Cassabois}}]{nguyen_optically_2012}%
  \BibitemOpen
  \bibfield  {author} {\bibinfo {author} {\bibfnamefont {H.~S.}\ \bibnamefont
  {Nguyen}}, \bibinfo {author} {\bibfnamefont {G.}~\bibnamefont {Sallen}},
  \bibinfo {author} {\bibfnamefont {C.}~\bibnamefont {Voisin}}, \bibinfo
  {author} {\bibfnamefont {P.}~\bibnamefont {Roussignol}}, \bibinfo {author}
  {\bibfnamefont {C.}~\bibnamefont {Diederichs}}, \ and\ \bibinfo {author}
  {\bibfnamefont {G.}~\bibnamefont {Cassabois}},\ }\href {\doibase
  10.1103/PhysRevLett.108.057401} {\bibfield  {journal} {\bibinfo  {journal}
  {Physical Review Letters}\ }\textbf {\bibinfo {volume} {108}},\ \bibinfo
  {pages} {057401} (\bibinfo {year} {2012})}\BibitemShut {NoStop}%
\bibitem [{\citenamefont {Chen}\ \emph {et~al.}(2016)\citenamefont {Chen},
  \citenamefont {Lander}, \citenamefont {Krowpman}, \citenamefont {Solomon},\
  and\ \citenamefont {Flagg}}]{chen_characterization_2016}%
  \BibitemOpen
  \bibfield  {author} {\bibinfo {author} {\bibfnamefont {D.}~\bibnamefont
  {Chen}}, \bibinfo {author} {\bibfnamefont {G.~R.}\ \bibnamefont {Lander}},
  \bibinfo {author} {\bibfnamefont {K.~S.}\ \bibnamefont {Krowpman}}, \bibinfo
  {author} {\bibfnamefont {G.~S.}\ \bibnamefont {Solomon}}, \ and\ \bibinfo
  {author} {\bibfnamefont {E.~B.}\ \bibnamefont {Flagg}},\ }\href {\doibase
  10.1103/PhysRevB.93.115307} {\bibfield  {journal} {\bibinfo  {journal}
  {Physical Review B}\ }\textbf {\bibinfo {volume} {93}},\ \bibinfo {pages}
  {115307} (\bibinfo {year} {2016})}\BibitemShut {NoStop}%
\bibitem [{\citenamefont {Aslam}\ \emph {et~al.}(2013)\citenamefont {Aslam},
  \citenamefont {Waldherr}, \citenamefont {Neumann}, \citenamefont {Jelezko},\
  and\ \citenamefont {Wrachtrup}}]{aslam_photo-induced_2013}%
  \BibitemOpen
  \bibfield  {author} {\bibinfo {author} {\bibfnamefont {N.}~\bibnamefont
  {Aslam}}, \bibinfo {author} {\bibfnamefont {G.}~\bibnamefont {Waldherr}},
  \bibinfo {author} {\bibfnamefont {P.}~\bibnamefont {Neumann}}, \bibinfo
  {author} {\bibfnamefont {F.}~\bibnamefont {Jelezko}}, \ and\ \bibinfo
  {author} {\bibfnamefont {J.}~\bibnamefont {Wrachtrup}},\ }\href {\doibase
  10.1088/1367-2630/15/1/013064} {\bibfield  {journal} {\bibinfo  {journal}
  {New Journal of Physics}\ }\textbf {\bibinfo {volume} {15}},\ \bibinfo
  {pages} {013064} (\bibinfo {year} {2013})}\BibitemShut {NoStop}%
\bibitem [{\citenamefont {Togan}\ \emph {et~al.}(2010)\citenamefont {Togan},
  \citenamefont {Chu}, \citenamefont {Trifonov}, \citenamefont {Jiang},
  \citenamefont {Maze}, \citenamefont {Childress}, \citenamefont {Dutt},
  \citenamefont {Sørensen}, \citenamefont {Hemmer}, \citenamefont {Zibrov},\
  and\ \citenamefont {Lukin}}]{togan_quantum_2010}%
  \BibitemOpen
  \bibfield  {author} {\bibinfo {author} {\bibfnamefont {E.}~\bibnamefont
  {Togan}}, \bibinfo {author} {\bibfnamefont {Y.}~\bibnamefont {Chu}}, \bibinfo
  {author} {\bibfnamefont {A.~S.}\ \bibnamefont {Trifonov}}, \bibinfo {author}
  {\bibfnamefont {L.}~\bibnamefont {Jiang}}, \bibinfo {author} {\bibfnamefont
  {J.}~\bibnamefont {Maze}}, \bibinfo {author} {\bibfnamefont {L.}~\bibnamefont
  {Childress}}, \bibinfo {author} {\bibfnamefont {M.~V.~G.}\ \bibnamefont
  {Dutt}}, \bibinfo {author} {\bibfnamefont {A.~S.}\ \bibnamefont {Sørensen}},
  \bibinfo {author} {\bibfnamefont {P.~R.}\ \bibnamefont {Hemmer}}, \bibinfo
  {author} {\bibfnamefont {A.~S.}\ \bibnamefont {Zibrov}}, \ and\ \bibinfo
  {author} {\bibfnamefont {M.~D.}\ \bibnamefont {Lukin}},\ }\href {\doibase
  10.1038/nature09256} {\bibfield  {journal} {\bibinfo  {journal} {Nature}\
  }\textbf {\bibinfo {volume} {466}},\ \bibinfo {pages} {730} (\bibinfo {year}
  {2010})}\BibitemShut {NoStop}%
\bibitem [{\citenamefont {De~Greve}\ \emph {et~al.}(2012)\citenamefont
  {De~Greve}, \citenamefont {Yu}, \citenamefont {McMahon}, \citenamefont
  {Pelc}, \citenamefont {Natarajan}, \citenamefont {Kim}, \citenamefont {Abe},
  \citenamefont {Maier}, \citenamefont {Schneider}, \citenamefont {Kamp},
  \citenamefont {Höfling}, \citenamefont {Hadfield}, \citenamefont {Forchel},
  \citenamefont {Fejer},\ and\ \citenamefont
  {Yamamoto}}]{de_greve_quantum-dot_2012}%
  \BibitemOpen
  \bibfield  {author} {\bibinfo {author} {\bibfnamefont {K.}~\bibnamefont
  {De~Greve}}, \bibinfo {author} {\bibfnamefont {L.}~\bibnamefont {Yu}},
  \bibinfo {author} {\bibfnamefont {P.~L.}\ \bibnamefont {McMahon}}, \bibinfo
  {author} {\bibfnamefont {J.~S.}\ \bibnamefont {Pelc}}, \bibinfo {author}
  {\bibfnamefont {C.~M.}\ \bibnamefont {Natarajan}}, \bibinfo {author}
  {\bibfnamefont {N.~Y.}\ \bibnamefont {Kim}}, \bibinfo {author} {\bibfnamefont
  {E.}~\bibnamefont {Abe}}, \bibinfo {author} {\bibfnamefont {S.}~\bibnamefont
  {Maier}}, \bibinfo {author} {\bibfnamefont {C.}~\bibnamefont {Schneider}},
  \bibinfo {author} {\bibfnamefont {M.}~\bibnamefont {Kamp}}, \bibinfo {author}
  {\bibfnamefont {S.}~\bibnamefont {Höfling}}, \bibinfo {author}
  {\bibfnamefont {R.~H.}\ \bibnamefont {Hadfield}}, \bibinfo {author}
  {\bibfnamefont {A.}~\bibnamefont {Forchel}}, \bibinfo {author} {\bibfnamefont
  {M.~M.}\ \bibnamefont {Fejer}}, \ and\ \bibinfo {author} {\bibfnamefont
  {Y.}~\bibnamefont {Yamamoto}},\ }\href {\doibase 10.1038/nature11577}
  {\bibfield  {journal} {\bibinfo  {journal} {Nature}\ }\textbf {\bibinfo
  {volume} {491}},\ \bibinfo {pages} {421} (\bibinfo {year}
  {2012})}\BibitemShut {NoStop}%
\end{thebibliography}%


\begin{thebibliography}{7}%
\makeatletter
\providecommand \@ifxundefined [1]{%
 \@ifx{#1\undefined}
}%
\providecommand \@ifnum [1]{%
 \ifnum #1\expandafter \@firstoftwo
 \else \expandafter \@secondoftwo
 \fi
}%
\providecommand \@ifx [1]{%
 \ifx #1\expandafter \@firstoftwo
 \else \expandafter \@secondoftwo
 \fi
}%
\providecommand \natexlab [1]{#1}%
\providecommand \enquote  [1]{``#1''}%
\providecommand \bibnamefont  [1]{#1}%
\providecommand \bibfnamefont [1]{#1}%
\providecommand \citenamefont [1]{#1}%
\providecommand \href@noop [0]{\@secondoftwo}%
\providecommand \href [0]{\begingroup \@sanitize@url \@href}%
\providecommand \@href[1]{\@@startlink{#1}\@@href}%
\providecommand \@@href[1]{\endgroup#1\@@endlink}%
\providecommand \@sanitize@url [0]{\catcode `\\12\catcode `\$12\catcode
  `\&12\catcode `\#12\catcode `\^12\catcode `\_12\catcode `\%12\relax}%
\providecommand \@@startlink[1]{}%
\providecommand \@@endlink[0]{}%
\providecommand \url  [0]{\begingroup\@sanitize@url \@url }%
\providecommand \@url [1]{\endgroup\@href {#1}{\urlprefix }}%
\providecommand \urlprefix  [0]{URL }%
\providecommand \Eprint [0]{\href }%
\providecommand \doibase [0]{http://dx.doi.org/}%
\providecommand \selectlanguage [0]{\@gobble}%
\providecommand \bibinfo  [0]{\@secondoftwo}%
\providecommand \bibfield  [0]{\@secondoftwo}%
\providecommand \translation [1]{[#1]}%
\providecommand \BibitemOpen [0]{}%
\providecommand \bibitemStop [0]{}%
\providecommand \bibitemNoStop [0]{.\EOS\space}%
\providecommand \EOS [0]{\spacefactor3000\relax}%
\providecommand \BibitemShut  [1]{\csname bibitem#1\endcsname}%
\let\auto@bib@innerbib\@empty
\bibitem [{\citenamefont {Sohn}\ \emph {et~al.}(2018)\citenamefont {Sohn},
  \citenamefont {Meesala}, \citenamefont {Pingault}, \citenamefont {Atikian},
  \citenamefont {Holzgrafe}, \citenamefont {Gündoğan}, \citenamefont
  {Stavrakas}, \citenamefont {Stanley}, \citenamefont {Sipahigil},
  \citenamefont {Choi}, \citenamefont {Zhang}, \citenamefont {Pacheco},
  \citenamefont {Abraham}, \citenamefont {Bielejec}, \citenamefont {Lukin},
  \citenamefont {Atatüre},\ and\ \citenamefont
  {Lončar}}]{sohn_controlling_2018}%
  \BibitemOpen
  \bibfield  {author} {\bibinfo {author} {\bibfnamefont {Y.-I.}\ \bibnamefont
  {Sohn}}, \bibinfo {author} {\bibfnamefont {S.}~\bibnamefont {Meesala}},
  \bibinfo {author} {\bibfnamefont {B.}~\bibnamefont {Pingault}}, \bibinfo
  {author} {\bibfnamefont {H.~A.}\ \bibnamefont {Atikian}}, \bibinfo {author}
  {\bibfnamefont {J.}~\bibnamefont {Holzgrafe}}, \bibinfo {author}
  {\bibfnamefont {M.}~\bibnamefont {Gündoğan}}, \bibinfo {author}
  {\bibfnamefont {C.}~\bibnamefont {Stavrakas}}, \bibinfo {author}
  {\bibfnamefont {M.~J.}\ \bibnamefont {Stanley}}, \bibinfo {author}
  {\bibfnamefont {A.}~\bibnamefont {Sipahigil}}, \bibinfo {author}
  {\bibfnamefont {J.}~\bibnamefont {Choi}}, \bibinfo {author} {\bibfnamefont
  {M.}~\bibnamefont {Zhang}}, \bibinfo {author} {\bibfnamefont {J.~L.}\
  \bibnamefont {Pacheco}}, \bibinfo {author} {\bibfnamefont {J.}~\bibnamefont
  {Abraham}}, \bibinfo {author} {\bibfnamefont {E.}~\bibnamefont {Bielejec}},
  \bibinfo {author} {\bibfnamefont {M.~D.}\ \bibnamefont {Lukin}}, \bibinfo
  {author} {\bibfnamefont {M.}~\bibnamefont {Atatüre}}, \ and\ \bibinfo
  {author} {\bibfnamefont {M.}~\bibnamefont {Lončar}},\ }\href {\doibase
  10.1038/s41467-018-04340-3} {\bibfield  {journal} {\bibinfo  {journal}
  {Nature Communications}\ }\textbf {\bibinfo {volume} {9}},\ \bibinfo {pages}
  {2012} (\bibinfo {year} {2018})}\BibitemShut {NoStop}%
\bibitem [{\citenamefont {Siyushev}\ \emph {et~al.}(2017)\citenamefont
  {Siyushev}, \citenamefont {Metsch}, \citenamefont {Ijaz}, \citenamefont
  {Binder}, \citenamefont {Bhaskar}, \citenamefont {Sukachev}, \citenamefont
  {Sipahigil}, \citenamefont {Evans}, \citenamefont {Nguyen}, \citenamefont
  {Lukin}, \citenamefont {Hemmer}, \citenamefont {Palyanov}, \citenamefont
  {Kupriyanov}, \citenamefont {Borzdov}, \citenamefont {Rogers},\ and\
  \citenamefont {Jelezko}}]{siyushev_optical_2017}%
  \BibitemOpen
  \bibfield  {author} {\bibinfo {author} {\bibfnamefont {P.}~\bibnamefont
  {Siyushev}}, \bibinfo {author} {\bibfnamefont {M.~H.}\ \bibnamefont
  {Metsch}}, \bibinfo {author} {\bibfnamefont {A.}~\bibnamefont {Ijaz}},
  \bibinfo {author} {\bibfnamefont {J.~M.}\ \bibnamefont {Binder}}, \bibinfo
  {author} {\bibfnamefont {M.~K.}\ \bibnamefont {Bhaskar}}, \bibinfo {author}
  {\bibfnamefont {D.~D.}\ \bibnamefont {Sukachev}}, \bibinfo {author}
  {\bibfnamefont {A.}~\bibnamefont {Sipahigil}}, \bibinfo {author}
  {\bibfnamefont {R.~E.}\ \bibnamefont {Evans}}, \bibinfo {author}
  {\bibfnamefont {C.~T.}\ \bibnamefont {Nguyen}}, \bibinfo {author}
  {\bibfnamefont {M.~D.}\ \bibnamefont {Lukin}}, \bibinfo {author}
  {\bibfnamefont {P.~R.}\ \bibnamefont {Hemmer}}, \bibinfo {author}
  {\bibfnamefont {Y.~N.}\ \bibnamefont {Palyanov}}, \bibinfo {author}
  {\bibfnamefont {I.~N.}\ \bibnamefont {Kupriyanov}}, \bibinfo {author}
  {\bibfnamefont {Y.~M.}\ \bibnamefont {Borzdov}}, \bibinfo {author}
  {\bibfnamefont {L.~J.}\ \bibnamefont {Rogers}}, \ and\ \bibinfo {author}
  {\bibfnamefont {F.}~\bibnamefont {Jelezko}},\ }\href {\doibase
  10.1103/PhysRevB.96.081201} {\bibfield  {journal} {\bibinfo  {journal}
  {Physical Review B}\ }\textbf {\bibinfo {volume} {96}},\ \bibinfo {pages}
  {081201} (\bibinfo {year} {2017})}\BibitemShut {NoStop}%
\bibitem [{\citenamefont {Loudon}(2000)}]{loudon_quantum_2000}%
  \BibitemOpen
  \bibfield  {author} {\bibinfo {author} {\bibfnamefont {R.}~\bibnamefont
  {Loudon}},\ }\href@noop {} {\emph {\bibinfo {title} {The {Quantum} {Theory}
  of {Light}}}},\ \bibinfo {edition} {3rd}\ ed.\ (\bibinfo  {publisher} {OUP
  Oxford University Press},\ \bibinfo {year} {2000})\BibitemShut {NoStop}%
\bibitem [{\citenamefont {Boldyrev}\ \emph {et~al.}(2018)\citenamefont
  {Boldyrev}, \citenamefont {Mavrin}, \citenamefont {Sherin},\ and\
  \citenamefont {Popova}}]{boldyrev_bright_2018}%
  \BibitemOpen
  \bibfield  {author} {\bibinfo {author} {\bibfnamefont {K.~N.}\ \bibnamefont
  {Boldyrev}}, \bibinfo {author} {\bibfnamefont {B.~N.}\ \bibnamefont
  {Mavrin}}, \bibinfo {author} {\bibfnamefont {P.~S.}\ \bibnamefont {Sherin}},
  \ and\ \bibinfo {author} {\bibfnamefont {M.~N.}\ \bibnamefont {Popova}},\
  }\href {\doibase 10.1016/j.jlumin.2017.07.031} {\bibfield  {journal}
  {\bibinfo  {journal} {Journal of Luminescence}\ }\bibinfo {series} {Special
  {Issue} in honor of {Professor} {Renata} {Reisfeld} for her outstanding
  contributions to luminescent inorganic glasses},\ \textbf {\bibinfo {volume}
  {193}},\ \bibinfo {pages} {119} (\bibinfo {year} {2018})}\BibitemShut
  {NoStop}%
\bibitem [{\citenamefont {Turukhin}\ \emph {et~al.}(1996)\citenamefont
  {Turukhin}, \citenamefont {Liu}, \citenamefont {Gorokhovsky}, \citenamefont
  {Alfano},\ and\ \citenamefont {Phillips}}]{turukhin_picosecond_1996}%
  \BibitemOpen
  \bibfield  {author} {\bibinfo {author} {\bibfnamefont {A.~V.}\ \bibnamefont
  {Turukhin}}, \bibinfo {author} {\bibfnamefont {C.-H.}\ \bibnamefont {Liu}},
  \bibinfo {author} {\bibfnamefont {A.~A.}\ \bibnamefont {Gorokhovsky}},
  \bibinfo {author} {\bibfnamefont {R.~R.}\ \bibnamefont {Alfano}}, \ and\
  \bibinfo {author} {\bibfnamefont {W.}~\bibnamefont {Phillips}},\ }\href
  {\doibase 10.1103/PhysRevB.54.16448} {\bibfield  {journal} {\bibinfo
  {journal} {Physical Review B}\ }\textbf {\bibinfo {volume} {54}},\ \bibinfo
  {pages} {16448} (\bibinfo {year} {1996})}\BibitemShut {NoStop}%
\bibitem [{\citenamefont {Neu}\ \emph {et~al.}(2012)\citenamefont {Neu},
  \citenamefont {Agio},\ and\ \citenamefont {Becher}}]{neu_photophysics_2012}%
  \BibitemOpen
  \bibfield  {author} {\bibinfo {author} {\bibfnamefont {E.}~\bibnamefont
  {Neu}}, \bibinfo {author} {\bibfnamefont {M.}~\bibnamefont {Agio}}, \ and\
  \bibinfo {author} {\bibfnamefont {C.}~\bibnamefont {Becher}},\ }\href
  {\doibase 10.1364/OE.20.019956} {\bibfield  {journal} {\bibinfo  {journal}
  {Optics Express}\ }\textbf {\bibinfo {volume} {20}},\ \bibinfo {pages}
  {19956} (\bibinfo {year} {2012})}\BibitemShut {NoStop}%
\bibitem [{\citenamefont {Thiering}\ and\ \citenamefont
  {Gali}(2018)}]{thiering_ab_2018}%
  \BibitemOpen
  \bibfield  {author} {\bibinfo {author} {\bibfnamefont {G.}~\bibnamefont
  {Thiering}}\ and\ \bibinfo {author} {\bibfnamefont {A.}~\bibnamefont
  {Gali}},\ }\href {\doibase 10.1103/PhysRevX.8.021063} {\bibfield  {journal}
  {\bibinfo  {journal} {Physical Review X}\ }\textbf {\bibinfo {volume} {8}},\
  \bibinfo {pages} {021063} (\bibinfo {year} {2018})}\BibitemShut {NoStop}%
\end{thebibliography}%
\end{document}


\title{Supplementary Material for\\ 
		\small ``Optical switching of resonance fluorescence from a single germanium vacancy color center in diamond''}
	
	\author{{Disheng Chen$^{1, 2}$, Mu Zhao$^1$, Yu Zhou$^1$, Johannes Froech$^3$, Carole Diederichs$^4$, Nikolay Zheludev$^{1,2,5}$, Igor Aharonovich$^3$, Wei-bo Gao$^{1, 2,}$\vspace{0.2cm}\\}
		{\small $^1$Division of Physics and Applied Physics, School of Physical and Mathematical Sciences, Nanyang Technological University, Singapore 637371, Singapore\\	
			$^2$The Photonics Institute and Centre for Disruptive Photonic Technologies, Nanyang Technological University, Singapore 637371, Singapore\\
			$^3$School of Mathematical and Physical Sciences, University of Technology Sydney, Ultimo, NSW, 2007, Australia\\
			$^4$MajuLab, International Joint Research Unit UMI 3654, CNRS, Université Côte d’Azur, \\ 
			Sorbonne Université, National University of Singapore, Nanyang Technological University, Singapore\\
			$^5$Optoelectronics Research Centre, University of Southampton, UK}}
	
	
	\maketitle
	
	\renewcommand\theequation{S\arabic{equation}}
	\renewcommand{\thefigure}{S\arabic{figure}}
	
	In this Supplementary Material, we provide further information on:\vspace{0.1cm}\\
	\hspace*{0.6cm}Section 1:  Sample preparation \\
	\hspace*{0.6cm}Section 2:  Supplementary data - lifetime, resonant \& non-resonant saturation power, $g^{(2)}(\tau)$, and AOM IRF \\
	\hspace*{0.6cm}Section 3:  PLE Analysis \& other Germanium vacancy centers \\
	\hspace*{0.6cm}Section 4:  Model - Rabi oscillation \& dark-state \\
	\hspace*{0.6cm}Section 5:  Rate analysis \& possible physical process \vspace{0.1cm}
	
	All the figures and equations in Supplementary Material are labeled with the prefix ``S'' to distinguish from those that appear in the body of the Letter.
	
	\begin{figure*}[b]
		\includegraphics{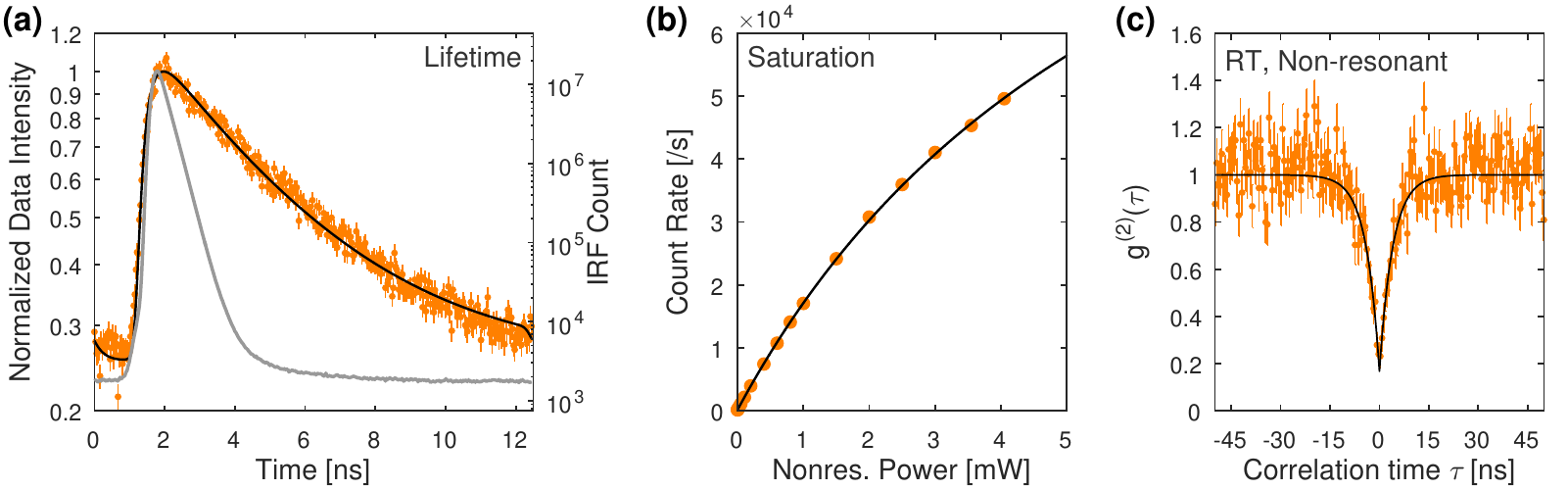}
		\caption{\label{fig:combo}  (a) Time-resolved PL of the GeV center under picosecond-pulse excitation, plotted in semi-logarithmic scale. The temperature is at 5 K. Orange dots are the raw data with the errorbar reflecting shot-noise. Grey curve is IRF of experimental setup. Black line is a fitting by convolving an exponential decay with the IRF. (b) Non-resonant power dependence of PL at 5 K under excitation at 532 nm. Orange dots are the raw data with the errorbar reflecting statistical fluctuation of the count rate over $\sim$2 mins. Solid purple curve is a fitting by using conventional saturation function. (c) 
		Room temperature second-order auto-correlation function $g^{(2)}(\tau)$ of the GeV center under non-resonant excitation (532 nm, $\sim$1 mW). Solid black line is a fitting with a single exponential decay.}
	\end{figure*}

	\section{SECTION 1: Sample Preparation}
	The substrate is an electronic grade diamond. Prior to all fabrication the sample was cleaned in hot Piranha acid ($150 ^\circ \text{C}/3:1~\text{H}_2\text{SO}_2:\text{H}_2\text{O}_2$) for 1 hr. After cleaning the sample was sputter coated with a Gold/Palladium film of 10 nm thickness for charge mitigation during processing. The micro-structures were fabricated using a FEI DB235, using a Ga+ ion beam with primary energy of 30 kV and a beam current of 7 nA. The fabricated solid immersion lens (SIL) structure has an approximate diameter of 5 $\mu$m and a height of 2.5 $\mu$m. After fabrication the Au/Pd film was removed using hot Aqua Regia ($100^\circ \text{C}/ 3:1 ~\text{HCl:HNO}_3$). Ge implantation was done with a charge state of 4, energy of 7 MeV, fluence of $5\times10^{10}$ cm$^{-2}$ at a current of 1.5 nA. The expected density of Germanium-vacancy (GeV) center is approximately 1$\sim$3 per SIL. After implantation the diamond was again cleaned in hot Piranha Acid. Subsequently the sample was annealed in high vacuum ($2\times10^{-6}$ mbar) at 900$^\circ$C for 2 hours.

	\begin{figure*}[b]
		\includegraphics{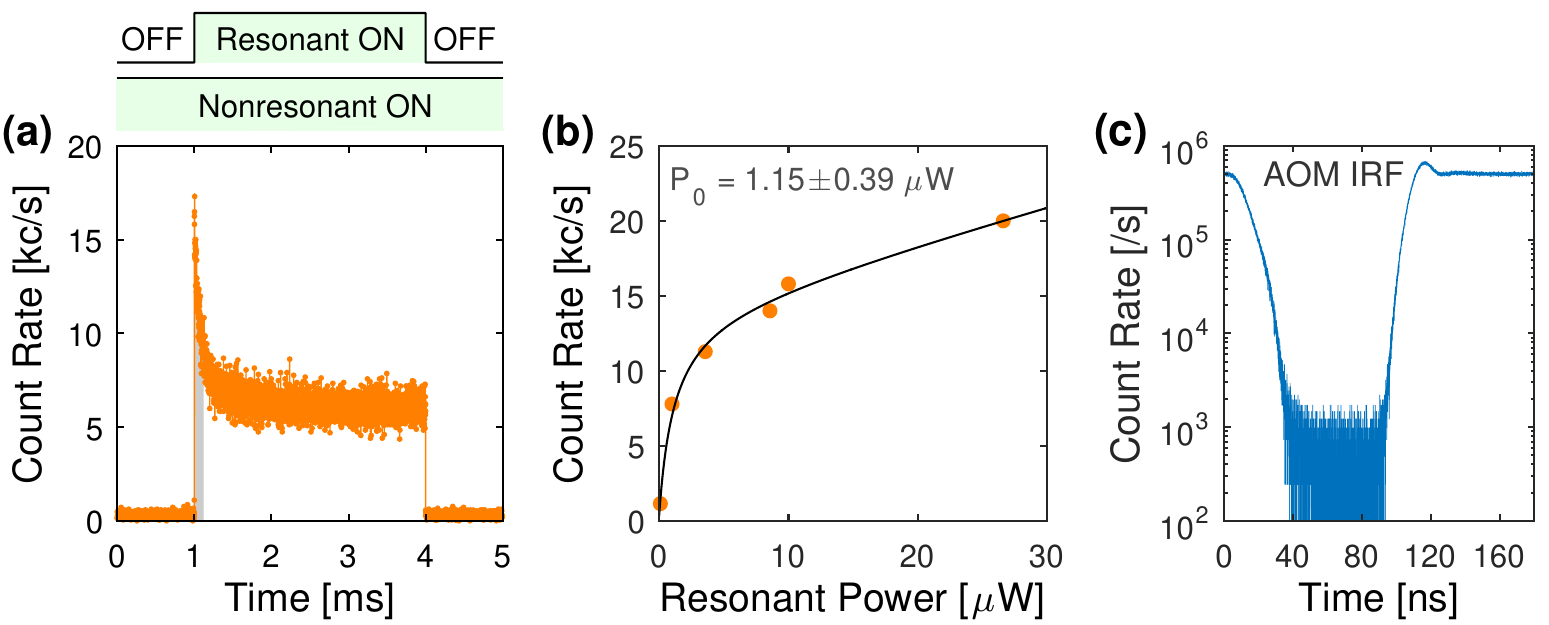}
		\caption{\label{fig:satu} (a) Time-dependent PL when modulating resonant laser while keeping non-resonant in cw mode. Shaded grey region corresponds to the time window within which the effect of resonant pumping is negligible. Top panel: modulation protocol. Resonant power: $8.7~\text{P}_0$. Non-resonant power: $7.4\times10^{-5}~\text{P}_1$.  (b) Saturation behavior of the PL. Orange dots are the extracted maximum intensity in (a). Black curve is a fitting by using the conventional saturation function plus a linear laser scattering background. (c) IRF of an AOM used for dynamics study, characterized at 532 nm. }
	\end{figure*}

	\section{SECTION 2: Supplementary Data}
	
	\noindent  \textbf{1. Lifetime T$_1$} \\
	\noindent\hspace*{0.35cm} Lifetime of the GeV center is determined by time-resolved PL measurement, as shown in Fig.~\ref{fig:combo}(a). A train of 512-nm 80-ps pulses at a repetition rate of 80 MHz (PiL051X) is focused onto the GeV center for excitation. Zero-phonon line (ZPL) emission of the emitter is collected by passing the light through a band-pass (600$\pm$7 nm) and a long-pass edge (600 nm) filter. By convolving an exponential decay with the measured instrument response function (IRF, grey curve) of the setup, lifetime of the GeV center is determined to be T$_1 = 3.57 \pm 0.03$ ns from the fitting.\vspace{0.1cm}
	
	
	\noindent \textbf{2. Non-resonant saturation power P$_1$} \\
	\noindent\hspace*{0.35cm} By fitting the power-dependent PL intensity with the conventional saturation function I$\propto$P/(P+P$_1$), the non-resonant saturation power (at 532 nm) of the GeV center is determined to be P$_1=6.8\pm0.2$ mW, as shown in Fig.~\ref{fig:combo}(b).\vspace{0.1cm}
	
	\noindent \textbf{3. Non-resonant $g^{(2)}(\tau)$} \\
	\noindent\hspace*{0.35cm} Figure~\ref{fig:combo}(c) shows the second-order correlation function of the GeV center, measured under non-resonant excitation at room temperature by using Hanbury Brown and Twiss (HBT) setup. The value of $g^{(2)}(0)=0.17\pm0.03$ unambiguously confirms the singleness of the emitter.  \vspace{0.1cm} 
	
	\noindent \textbf{4. Resonant saturation power P$_0$} \\
	\noindent\hspace*{0.35cm} 	To pinpoint the resonant saturation power P$_0$, we employ a pulsed measurement scheme to recover the saturation behavior of PL, as shown in Fig.~\ref{fig:satu}. In this experiment, the resonant beam is modulated by passing through an acousto-optic modulator (AOM), while the non-resonant beam is in cw mode and keeps at a constant power. The dwell time between two consecutive resonant pulses is chosen to be a few milli-seconds
	such that the GeV center can always relax back to the same equilibrium state for all different resonant powers used. Only the PL intensity within the transient time window upon the excitation [as indicated by the shaded grey area in Fig.~\ref{fig:satu}(a)] is extracted for data analysis. By fitting the data with I$_\text{PL} = A\times \text{P}/(\text{P}+\text{P}_0)+B\times P$ with the second term describing the laser scattering, we obtain a saturation power of P$_0=1.15\pm0.3$  $\mu$W, as shown in Fig.~\ref{fig:satu}(b). \vspace{0.1cm} 
	
	\noindent \textbf{5. IRF of acousto-optic modulator} \\
	\noindent\hspace*{0.35cm} Figure~\ref{fig:satu}(c) shows the IRF of the acousto-optic modulator (AOM) used in dynamics measurement. The response time is as short as 10 ns, which is three orders of magnitude faster than the dynamics being characterized. \vspace{0.1cm}

	\begin{figure*}[b]
		\includegraphics{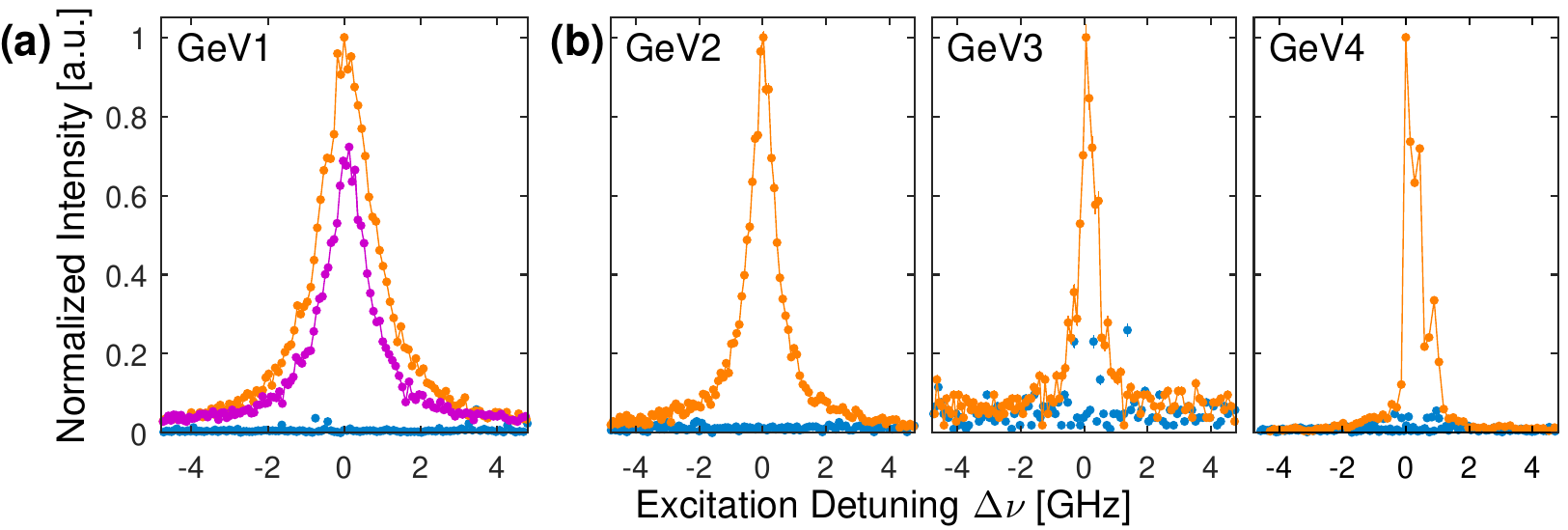}
		\caption{\label{fig:GeVs} (a) Normalized PLE scans of GeV1 (original GeV center used in the Letter) collected at different gating wavelength, i.e., 532 nm (orange) and 405 nm (purple). Blue trace is gating off. Zero detuning: 497.7541 THz; normalization constant: 10530 Hz. (b) Optical switching observed on other three GeV centers. For each emitter, two PLE spectra (normalized) are present with gating on (orange) or off (blue). The gating laser is at 532 nm. Zero detuning corresponds to 497.7686 THz, 497.7585 THz, and 497.9935 THz for GeV2, GeV3, and GeV4, respectively; and normalization constant is 4490 Hz, 1040 Hz, 9820 Hz, and 10530 Hz, respectively. All spectra in (a) and (b) are measured at the same resonant power of 300 nW. Non-resonant power is 1.3 $\mu$W for both 532 nm and 405 nm.  }
	\end{figure*}

	\section{SECTION 3: PLE Analysis \& other Germanium vacancy centers}	
	The disappearance of extra peaks around transition D in Fig.~2(a) of the Letter is possibly caused by the non-uniform strain across the SIL. GeV centers located at different positions in the SIL would experience different strains, which is expected to have little influence on transition C but significant impact on transition D based on the previous studies on silicon-vacancy center (SiV) \cite{sohn_controlling_2018}. Non-zero Eg-symmetric strains (perpendicular to the C3 symmetry axis of SiV center) would enhance the fine structure splittings in both ground- and excited-manifold by a similar magnitude, i.e., the absolute shift of each energy level is almost the same. This leads to a significant variation in transition energies for A and D while little for B and C. Considering the almost identical electronic structure of GeV center to SiV center, the strain is expected to have an similar effect on GeV center. Thereby, D lines of the nearby GeV centers must be shifted out of the measurement window thanks to the different strains experienced by centers. 
	
	Here, we provide a detailed explanation on broadening of PLE spectrum for low non-resonant power, as shown in Fig.~3(g) of the Letter. The cause is the detuning-dependent shelving efficiency of the system, which depends both shelving and deshelving processes. When the non-resonant power is low, the resonant-induced shelving rate is comparable to the non-resonant-induced shelving and deshevling rates. Thus, a change in any of these rates can impact the overall shelving efficiency significantly, which is what happens when sweeping the resonant laser. Specifically, detuning decreases the shelving efficiency by reducing the available population in the excited state. RF intensity is thus less influenced or suppressed for a larger detuning. This leads to a distortion of PLE line shape as if it is flattened, i.e., broadened. The situation is different, however, when a strong non-resonant beam involves. In this scenario, the non-resonant laser induced shelving and deshelving rates are enhanced and can dominate the population dynamics in the system. And the variation of resonant-induced shelving rate is not able to affect the shelving efficiency due to the small magnitude. Consequentially, The excitation linewidth can maintain its intrinsic line shape (intrinsic line width) without experiencing any significant distortion as if no shelving effect exists. 
	
	Figure~\ref{fig:GeVs}(a) shows PLE scans of the original GeV center (GeV1) investigated in the Letter by gating at different wavelengths, i.e., 405 nm (purple trace) and 532 nm (orange trace). Since we keep the non-resonant beam at the same power of 1.3 $\mu$W for both scans, the number of photons in 405 nm beam is 76\% of those in 532 nm beam. This qualitatively explains the shrink of PLE spectra for 405 nm. In fact, the decrease of the intensity is very close to 76\%, indicating that the gating efficency of 405 nm light is almost the same as 532 nm light. Therefore, the non-resonant beam must interact with the system incoherently to generate the shelving and deshelving processes.   
	
	Optical switching/gaiting of resonance fluorescence is not limited to GeV1, but a ubiquitous phenomenon, observable on almost all GeV centers we investigated so far from different samples (more than 20 centers). 
	Figure~\ref{fig:GeVs}(b) shows three examples of them by collecting two PLE spectra for each with gating on and off. For all three emitters, the resonance fluorescence is quenched when the gating laser is off (blue trace), and the RF is recovered when introducing a small amount of non-resonant light (orange trace). In fact, a similar blinking of RF has been reported in a HPHT diamond, where the GeV centers are incorporated during the growth in a Mg-Ge-C system \cite{siyushev_optical_2017}.

	\section{SECTION 4: Model - Rabi oscillation \& dark-state}	
	
	For a 2-level system under resonant pumping, the population in excited state displays an oscillatory behavior, i.e., Rabi oscillation, resulting from the competition between resonant excitation (absorption) and stimulated emission. In reality, this coherent process is constantly disturbed/interrupted by various dephasing mechanism, such as spontaneous decay and phonon interaction, causing a damping of the oscillation over a period defined by the coherence time of the excited state, i.e., T$_2$. Experimentally, we take advantage of the spontaneous decay to map the oscillatory population into the PL intensity by accurately timing each excitation pulse generated by an electro-optic modulator (EOM). Note that the AOM is not fast enough to resolve the oscillation. We employ a simple 2-level model formulated in semiclassical picture for theoretical description of Rabi oscillation. The time evolution of the system follows the master equation in Lindblad form \cite{loudon_quantum_2000}, which has taken both spontaneous decay and pure dephasing into account
	\begin{align}
	\label{eqn:2lv}
	&\frac{d}{dt} \left(\begin{array}{c}
	\rho_\text{G} \\ \rho_\text{E} \\ \rho_\text{GE} \\ \rho_\text{EG}
	\end{array}\right) =
	\left(
	\begin{array}{cccc}
	0 & \Gamma_\text{sp} & i\Omega/2 & -i\Omega/2 \\
	0 & -\Gamma_\text{sp} & -i\Omega/2 & i\Omega/2 \\
	i\Omega/2 & -i\Omega/2 & -1/T_2 & 0 \\
	-i\Omega/2 & i\Omega/2 & 0 & -1/T_2
	\end{array}\right) \left( 
	\begin{array}{c}
	\rho_\text{G}\\
	\rho_\text{E}\\
	\rho_\text{GE}\\
	\rho_\text{EG}
	\end{array}\right)
	\end{align}
	where $\rho_\text{G}$, $\rho_\text{E}$ are the time-dependent population in ground and excited state, $\rho_\text{GE}$, $\rho_\text{GE}$ are the coherence between G and E, $\Omega$ is the resonant Rabi frequency, $\Gamma_\text{sp}$ is the spontaneous decay rate, and $\text{T}_2$ is the coherence time of excited state. The time-dependent PL intensity is proportional to $\rho_\text{E}$(t)$\Gamma_\text{sp}$ by a constant that relies on the efficiency of the entire system, including the quantum yield of GeV center \cite{boldyrev_bright_2018}. By solving Eqn.~\ref{eqn:2lv} numerically, we fit multiple oscillations taken at different excitation powers simultaneously to pinpoint a coherence time of T$_2=366\pm20$ ps. The Rabi frequencies obtained in the fitting follows linearly over the sqaure root of the excitation power as shown in the inset of Fig.~2(d) of the Letter. The success of 2-level modeling indicates that within the short period of measurement pulse, the dark-state related dynamics has a negligible effect on the system.   
	
	The PL intensity (under cw excitation) of the GeV center can be formulated by using the steady-state solution of Eqn.~\ref{eqn:2lv}
	\begin{equation}
	\label{eqn:2levelpop}
	\rho_\text{E}^\text{2-level} = \frac{1}{2} \frac{\Omega^2\gamma/\Gamma_\text{sp}}{\Delta\omega^2+\gamma^2+\Omega^2\gamma/\Gamma_\text{sp}}
	\end{equation}
	where $\Delta\omega$ is the excitation detuning, and $\gamma=1/\text{T}_2$. By comparing Eqn.~\ref{eqn:2levelpop} to the conventional saturation formula of $\rho_\text{E}\propto \text{P}/(\text{P}+\text{P}_0)$ where P$_0$ is resonant saturation power, we obtain the relationship between excitation power P and Rabi frequency $\Omega$ as
	\begin{equation}
	\label{eqn:Rabi}
	\Omega = \sqrt{\gamma\Gamma_\text{sp}\frac{\text{P}}{\text{P}_0}}
	\end{equation}
	
	For the GeV center investigated here, the presence of dark state and channel $k_\text{ED}$ necessitates the need of 3-level system for a complete description of the dynamics. Again, we formulate the model in semiclassical picture to incorporate the effect of stimulated emission. Since no other transitions are coherently pumped except transition G-E, Rabi frequencies connecting the other combinations of states must vanish. Thus, the block of evolution matrix for elements $\rho_\text{ED}$, $\rho_\text{GD}$, $\rho_\text{DG}$ and their complex conjugates are completely diagonalized, and is irrelevant to the solution of $\rho_\text{E}$. So we drop these terms in our formula and obtains a density matrix with five elements as shown in Eqn.(1) of the Letter. Its Steady-state solution gives
	\begin{equation}
	\label{eqn:popuE}
	\rho_\text{E}^{\text{3-level}} = \frac{1}{2} \frac{k_\text{DG}}{k_\text{DG}+k_\text{GD}}
	\left[\frac{(\Delta\omega^2+\gamma^2)(k_\text{ED}+\Gamma_\text{sp})} {\gamma\Omega^2} + 
	\frac{1}{2}\left(1+\frac{k_\text{ED}+k_\text{DG}}{k_\text{DG}+k_\text{GD}}\right)\right]^{-1}
	\end{equation} 
	where $k_\text{ED}$, $k_\text{GD}$, and $k_\text{DG}$ are the population transfer rates from excited to dark, ground to dark, and dark to ground, respectively. In fact, Eqn.~\ref{eqn:2levelpop} is a special case of Eqn.~\ref{eqn:popuE} by setting rates $k_\text{ED}$ and $k_\text{GD}$ to zero. The resonant-power-dependence of rates ($k_{\alpha\beta}$) are responsible for the drop of RF intensity for high excitation power in Fig.3(d) of the Letter. However, lack of knowledge on these power dependences prevents us from analyzing the data by using Eqn.~\ref{eqn:popuE}. 
	
	Equation~\ref{eqn:popuE} also defines the full-width half-maximum (FWHM) of an excitation spectrum, as given by Eqn.~(3) in the Letter. Again, without knowing the power dependences of $k_{\alpha\beta}$, it is impossible to use Eqn.~(3) in the Letter to examine the linewidth. Nevertheless, we can still extract the coherence time T$_2$ from the data by extrapolating the linewidth to zero excitation power, as discussed in the Letter.

	\section{SECTION 5: Rate analysis \& possible physical process}	

	In the dynamics study as shown in Fig.~4 of the Letter, several parameters in Eqn.~(1) are fixed to constants to validate the fitting. Some constants are determined by separate measurements, for example, spontaneous decay rate $\Gamma_\text{sp}=1/\text{T}_1 = 280$ MHz via lifetime measurement, coherence time T$_2 = 316$ ps through linewidth analysis; and Rabi frequency $\Omega$ by using Eqn.~\ref{eqn:Rabi} with the measured excitation power. At a glance, T$_2$ should not be a constant, because dark-state introduces an extra channel for dephasing whose rate depends on both resonant and non-resonant powers, as shown in Fig.~4(d) and (g). However, the dark-state dynamics is too slow to have a noticeable impact on the dephasing rate 1/T$_2$, which is already several orders of magnitude faster than $k_{\alpha\beta}$, even at zero excitation power.
	
	In addition, the overall efficiency $\eta$ in Eqn.~(2) is fixed to $9\times10^{-5}$ through the entire fitting procedure. In principle, $\eta$ is determined by two factors, i.e, $\eta=\eta_D\times\eta_Q$, where $\eta_D$ is the overall detection efficiency of the entire experimental setup, and $\eta_Q$ is the quantum yield of the GeV center. For our system, $\eta_D$ is found to be $\sim$0.3\% by considering collection efficiency of the objective, fiber coupling efficiency, detector efficiency, beam-splitter ratio, optic loss, phonon-side band (PSB) branching ratio. Quantum yield $\eta_Q$ is defined as the probability for a photon emission upon a transition from excited state to ground state, which is found to be 3\% based on the absorption measurement by Boldyrev et al \cite{boldyrev_bright_2018}. This quantum yield is close to the value measured for SiV- center \cite{turukhin_picosecond_1996, neu_photophysics_2012}.  
	
	For the modulation experiment shown in Fig.~4(c) of the Letter, resonant laser is in CW mode at a constant power level. Since $k_\text{ED}$ is mainly affected by resonant pumping, this rate is kept as a shared parameter for different non-resonant powers. In addition, rates $k_\text{GD}^\text{off}$ and $k_\text{DG}^\text{off}$ are also the same for different runs since the experimental conditions of off-period are identical for different runs. Therefore, we employ a global fitting method to firstly determine these three rates as following: $k_\text{ED}=173 \pm 16$ Hz, $k_\text{GD}^\text{off} = 92 \pm 28$ Hz, and $k_\text{DG}^\text{off} = 4 \pm 3$ Hz.
	
\begin{figure*}[b]
		\includegraphics{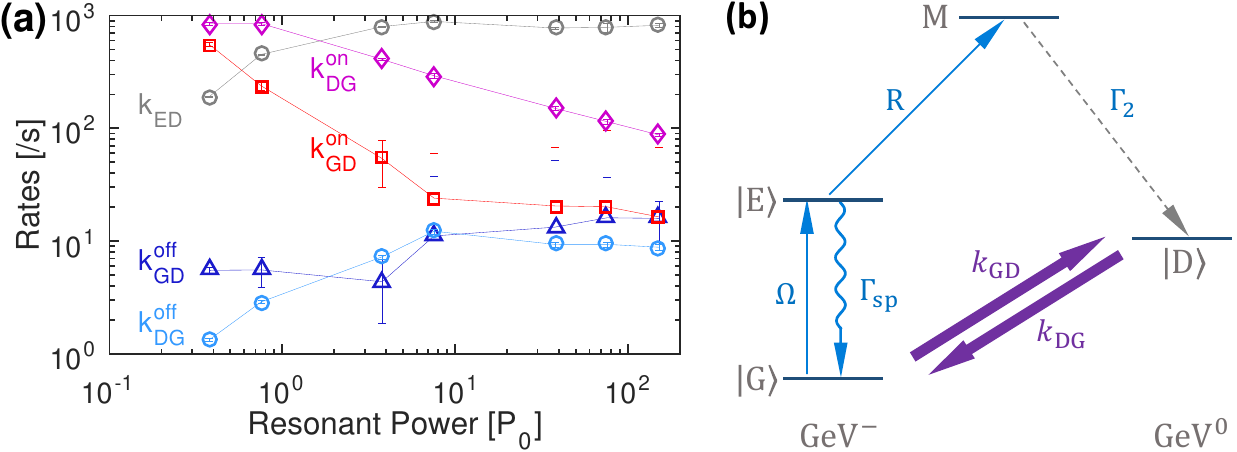}
		\caption{\label{fig:rates} Possible physical mechanism giving rise to the gating and shelving. State G, E, D amd M represents the ground, excited, dark and meta-stable state of GeV center, respectively. k$_\text{GD}$ is the transition rate from the ground state of GeV$^-$ to dark state GeV$^0$ via losing an electron, and k$_\text{DG}$ is the opposite process via losing an hole. $\Omega$ is Rabi frequency. $\Gamma_\text{sp}$ is the spontaneous decay rate of the excited state. R represents a population pumping channel from the excited state to the metastable state M, which is active with 602 nm laser. The population residing in M would relax to dark state D via a non-radiative channel at rate $\Gamma_\text{2}$. }
	\end{figure*}

	Following the same logic, when varying the resonant power as shown in Fig.~4(f) of the Letter, $k_\text{ED}$ should be the same for on and off period for each run, i.e., $k_\text{ED}^\text{off}=k_\text{ED}^\text{on}$. Figure~\ref{fig:rates}(a) shows all the extracted rates for different resonant powers, including those shown in Fig.~4(g). Rates $k_\text{GD}^\text{off}$ and $k_\text{DG}^\text{off}$ shows a slight variation over the resonant power implying that the resonant laser could influence the shelving (GD) and deshelving (DG) channels on its own.

	Figure~\ref{fig:rates}(b) shows one possible physical picture that underpins the observed shelving and deshelving phenomena. As mentioned in the Letter, the dark state (D) of the GeV center is possibly a GeV center in different charge states depending on the local Fermi level \cite{thiering_ab_2018}. Here, we temporarily assign it to be the neutral state (GeV$^0$), which has a different energy structure with respect to GeV$^-$ \cite{thiering_ab_2018}. The RF quenches when the GeV center resides in this neutral state; and the RF would recover if the charge state of GeV center returns back to negative. This explains the PL intermittency as observed in the inset of Fig.~2(c) of the Letter. 
	
	It is possible to stabilize the resonance fluorescence by employing a weak non-resonant laser. This light produces a free charge carrier bath around the GeV center by interacting with the local defects or impurities. These charge carriers modify the local Fermi level, and influence the charge dynamics of the emitter. Generally, stronger non-resonant power corresponds to higher charge-carrier density, thus faster transition rates k$_\text{GD}$ and k$_\text{DG}$, as indicated by the purple arrows in Fig.~\ref{fig:rates}(b). Resonant laser, on the other hand, can shelve the population of the system into dark-state D via a meta-stable state M, as shown in Fig.~\ref{fig:rates}(b). Pumping rate R is power-dependent, which is responsible for the enhancement of $k_\text{ED}$; while the non-radiative relaxation from M to D is power-independent, which accounts for the saturation of rate $k_\text{ED}$ at large excitation power as shown in Fig.~\ref{fig:rates}(a).
	
	When the GeV$^-$ center is solely pumped by the resonant laser, most population is shelved into the dark state,  leading to a quenching of RF. However, the background doping of the sample provides the needed charge carriers to enable the emitter to jump back to GeV$^-$ state via $k_\text{DG}$, which generates a short burst of RF as observed in Fig.~2(b) of the Letter. In addition, resonant laser can also interact with the local defects or impurities to produce empty charge traps that can effectively trap the free charge carriers introduced by the non-resonant laser. This explains the decrease of rates k$_\text{GD}$ and k$_\text{DG}$ in Fig.~4(g). 
	

	\bibliography{./GeV_blinking_supp_v2}